\documentclass{article}

\usepackage{arxiv}

\usepackage[]{geometry}
\usepackage[T1]{fontenc}
\usepackage[utf8]{inputenc}
\usepackage{lmodern}
\usepackage{microtype}
\usepackage{amsmath,amssymb,bm,mathtools}
\usepackage{graphicx}
\usepackage{booktabs,longtable,tabularx,array,multirow,calc}
\usepackage{caption}
\usepackage{pdflscape}
\usepackage{enumitem}
\usepackage{xcolor}
\usepackage[round,authoryear]{natbib}
\usepackage{xurl}
\usepackage[hidelinks]{hyperref}
\usepackage{etoolbox}

\newlength{\manuscripttablewidth}
\newcommand{\manuscripttablefont}{\footnotesize}
\AtBeginEnvironment{longtable}{\manuscripttablefont}

\graphicspath{ {./images/} }

\title{Variable-Horizon Workforce Demand Forecasting with an Aggregate Demand Constraint for Construction Workforce Planning}

\author{
 Hanbyeol Park \\
  Department of Industrial Engineering\\
  Pusan National University\\
  Pusan, Republic of Korea\\
  \texttt{pb104@pusan.ac.kr} \\
  \And
 Jaehyeon Heo \\
  Department of Industrial Engineering\\
  Pusan National University\\
  Pusan, Republic of Korea\\
  \texttt{jaehyeonheo07@gmail.com} \\
  \And
 Taekhyun Park \\
  Department of Industrial Engineering\\
  Pusan National University\\
  Pusan, Republic of Korea\\
  \texttt{pthpark1@pusan.ac.kr} \\
  \And
 Minseong Kim \\
  Smart Yard Research Center\\
  Samsung Heavy Industries Co.,Ltd\\
  Geoje, Republic of Korea\\
  \texttt{minsy.kim@samsung.com} \\
  \And
 Hyerim Bae \\
  Department of Industrial Engineering\\
  Pusan National University\\
  Pusan, Republic of Korea\\
  \texttt{hrbae@pusan.ac.kr} \\
}

\begin{document}
\maketitle
\begin{abstract}
Workforce planning is a recurring operational decision during construction projects that requires accurate forecasts of the future workforce demand for individual tasks. However, in practice, tasks have different completion dates, resulting in variable forecast horizons. In addition, the sum of the predicted daily workforce demands must equal the total workforce allocation specified in advance. Most existing machine-learning (ML)-based forecasting models assume fixed-length outputs and do not explicitly impose an aggregate demand constraint, making them unsuitable for these operational requirements. To address this problem, this study proposes constraint-preserving residual allocation forecasting (CP-RAF). The CP-RAF represents an observed workforce demand time series as a coefficient vector and retrieves completed tasks with similar temporal shapes. Then, it estimates the allocation profile over the remaining task duration using similarity–weight averaging. The predefined remaining workforce demand for each task was distributed according to the estimated profile, and the forecast horizon was adjusted while retaining profile characteristics. This procedure accommodates variable forecast horizons while preserving the aggregate demand constraints. CP-RAF was evaluated using workforce demand field data. The results showed that CP-RAF outperformed eight baseline models in medium- and long-horizon fixed-length forecasting and maintained low forecast errors under variable-length forecasting. By directly incorporating operational constraints into the forecasting procedure, the proposed method provides a framework suitable for workforce allocation in construction practices.
\end{abstract}

\section{Introduction}
\label{sec:sec1}

Workforce forecasting has been applied across a wide range of industries, including healthcare \citep{AzimiNayebi2018Nursing,Chung2010Nursing}, construction \citep{Cao2024Workforce}, and aviation \citep{Ye2016GreyForecast}. This is a central decision problem because it directly affects operational efficiency, cost control, and schedule adherence \citep{Micheli2023WorkforcePlanning}. Construction projects are characterized by substantial heterogeneity because each project has distinct requirements and may be at a different stage of execution \citep{Elkholosy2024DataMining}. Therefore, workforce demand forecasting should account for differences in project characteristics, and project-level workforce demand forecasts can directly support resource allocation and schedule management \citep{Wong2008Multivariate}. Accordingly, effective workforce planning and related operational decisions require accurate forecasting of future workforce demands \citep{Chan2006ConstructionSkills}.

Workforce forecasting methods can be classified into qualitative and quantitative approaches \citep{Ogungbire2025Workforce}. Qualitative approaches rely on structured expert judgments, such as the Delphi method \citep{AbLatif2016Delphi}. Quantitative approaches have traditionally employed statistical methods, such as AutoRegressive Integrated Moving Average (ARIMA), Vector Error Correction (VEC), and regression models \citep{Agarwal2013Model,Wong2005BoxJenkins,Wong2011ManpowerDemand}. These methods are interpretable and relatively straightforward to implement, and their standard formulations often rely on pre-specified linear relationships and temporal structures. Consequently, they may have difficulty in capturing nonlinear temporal dependencies and complex interactions among predictors unless such relationships are explicitly specified \citep{Ogungbire2025Workforce}. Machine learning (ML) methods, including long short-term memory (LSTM) networks, tree-based ensembles, and nonlinear methods, have recently been introduced to address these limitations \citep{Cao2024Workforce,Eichenseer2025DeliveryPositions,Ogungbire2025Workforce}. For example, tree-based ensemble models with neural time-series models have been used to forecast daily delivery positions for workforce planning in logistics operations \citep{Eichenseer2025DeliveryPositions}. Random forest (RF) regression has demonstrated strong performance in forecasting the project-level workforce demands of construction engineers and inspectors at state transportation agencies \citep{Ogungbire2025Workforce}. Extreme gradient boosting (XGBoost) has also produced accurate forecasts of maintenance workforce demand in the Hong Kong construction industry \citep{Cao2024Workforce}.

\begin{figure}[htbp]
	\centering
	\includegraphics[width=\linewidth]{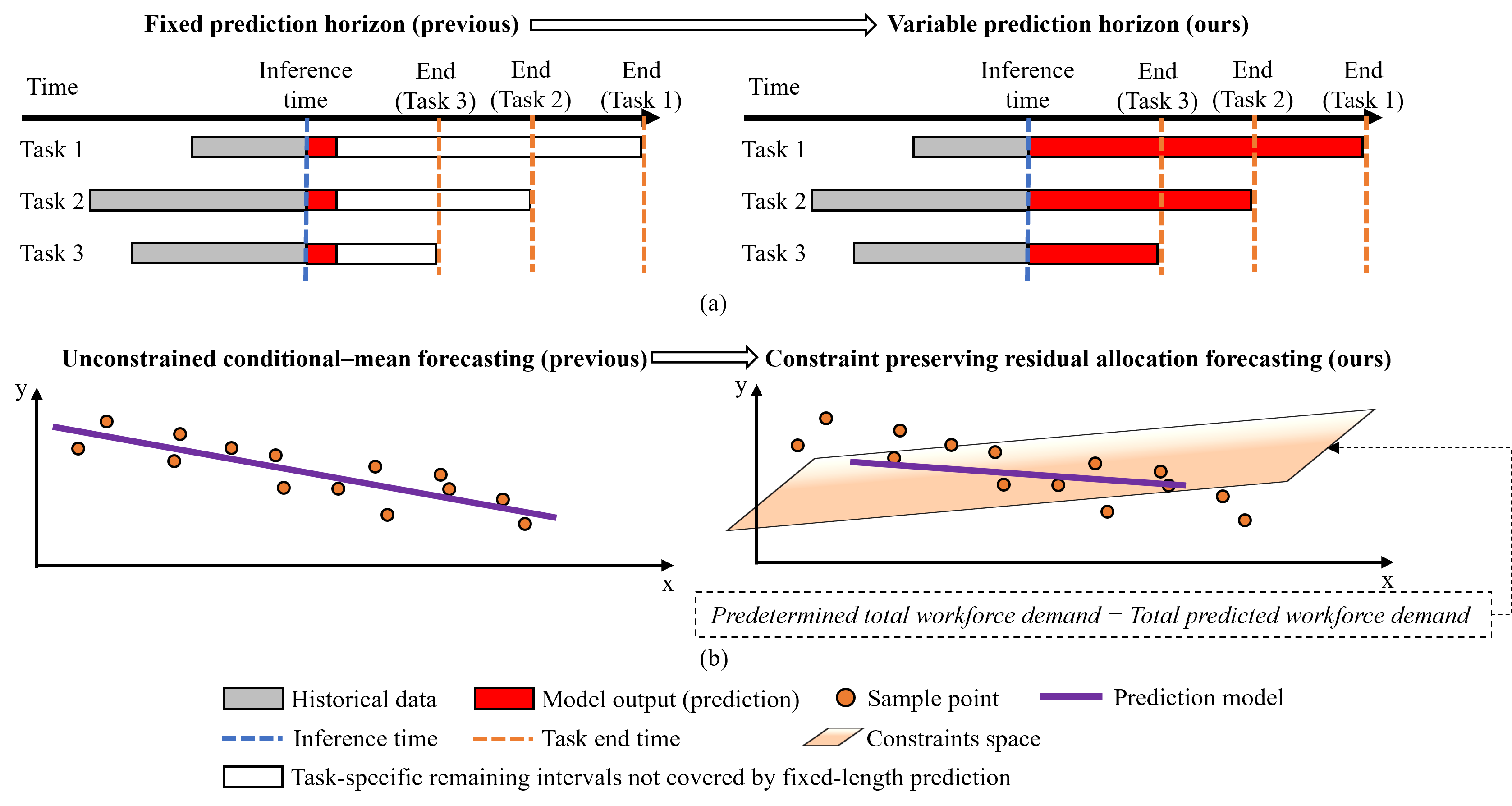}
	\caption{Differences between conventional forecasting approaches and the proposed approach. (a) Forecast-horizon perspective: fixed horizons in conventional studies and variable horizons in the proposed study. (b) Constraint perspective: conditional–mean function fitting in conventional studies and function fitting within the feasible constraint set in the proposed study.}
	\label{fig:fig1}
\end{figure}

Despite these results, the existing models remain limited as decision-support tools for construction site managers. Managers must determine how a pre-specified workforce should be distributed over the remaining duration of a task. This decision required two additional considerations. First, the completion date of a task depends on its current progress and the time at which the forecast is made. Therefore, forecasting models must accommodate different horizons. Second, the sum of the predicted workforce demands over the remaining task duration must equal the aggregate workforce allocation specified in advance. Most existing forecasting models assume fixed-length outputs and use unconstrained regressions without restricting the sum of their predictions. Therefore, they do not adequately support workforce-allocation decisions required in practice. These two requirements are illustrated in Figure \ref{fig:fig1}.

To address these limitations, this study proposes a method that represents the observed workforce demand time series of each task as a coefficient vector, identifies tasks with similar temporal shapes through clustering and retrieval, and estimates future workforce allocation profiles using similarity–weighted averaging. The final daily workforce demand is then obtained by distributing the predefined aggregate workforce requirements for each task according to the estimated profile. This formulation accommodates different forecast horizons and ensures that the sum of forecasts equals the predefined aggregate demand.

The main contributions of this study are as follows:

\begin{itemize}

    \item The variable-horizon forecasting problem arising from differences in task completion dates was formulated as a residual-demand allocation problem, and a forecasting framework applicable to variable-length horizons was developed.
    
    \item A coefficient-based representation links the temporal shape of the observed segment to the retrieval of reference tasks, thereby providing a means of estimating the workforce allocation profile over the remaining task duration.

\end{itemize}

The proposed method is evaluated using field data collected from a shipyard in Geoje, Republic of Korea. The experiments comprised two settings: (i) short, fixed-length forecasting with a maximum horizon of 15, and (ii) task-specific variable-length forecasting with a maximum horizon of 291. In the fixed-length experiments, the proposed model reduced the MAE by 1.416, 3.363, 3.081, and 5.188 at 5-, 7-, 10-, and 15-day horizons, respectively. Diebold–Mariano (DM) tests \citep{Diebold2012PredictiveAccuracy} were also conducted for 40 combinations of eight baseline models and five fixed forecast horizons. The improvement achieved by the proposed model was statistically significant for 30 comparisons. In the variable-length setting, the proposed model allowed the forecast horizon to vary across tasks and covered more forecast time points than the fixed-length setting yet produced lower forecast errors than the baseline models evaluated in the fixed-length experiments. These results indicate that the proposed method remains stable not only for short-term forecasts but also for medium- and long-term sequence forecasting. Its stability over long horizons is particularly relevant to workforce planning and operational decision-making in production environments, where forecasts must support decisions over the entire remaining duration of a task.

The remainder of this paper is organized as follows: Section \ref{sec:sec2} reviews the related literature and defines the research gaps addressed in this study. Section \ref{sec:sec3} presents the proposed methodology and Section \ref{sec:sec4} analyzes the experimental results. Finally, Section \ref{sec:sec5} concludes the paper and outlines directions for future research.

\section{Related works}
\label{sec:sec2}

Research on workforce demand forecasting in construction and logistics can be broadly classified into traditional statistical regression methods and recent ML approaches. This section reviews prior studies that have focused on the forms of their forecast outputs.

Conventional regression-based approaches offer clear model structures and straightforward interpretations. \citep{Wong2005BoxJenkins} applied Box–Jenkins ARIMA time-series models to forecast employment, productivity, unemployment, employment rates, and wages in the Hong Kong construction labor market. \citep{Wong2011ManpowerDemand} subsequently compared the Box–Jenkins ARIMA, multiple log-linear regression (LR), and VEC models to analyze the structural relationship between macroeconomic variables and construction employment. \citep{Agarwal2013Model} developed a cost-based LR to estimate the workforce requirements for real estate construction projects. \citep{Alqatawna2023Forecasting} forecast logistics order volumes to determine the workforce required by a company. Their seasonal ARIMA with exogenous regressor (SARIMAX) model incorporates exogenous variables, such as holidays and sales seasons, and improves forecast performance. Although these models are structurally transparent and readily interpretable, their standard formulations impose linear conditional mean structures, and are therefore inadequate for representing complex nonlinear temporal relationships \citep{SantosJunior2019HybridARIMA,Zhang2003HybridARIMA}. More importantly for the present study, the cited applications were developed for aggregate labor market series, project-level workforce totals, or company-level order volumes. They did not explicitly formulate daily task-level forecasts with horizons determined by task-specific completion dates \citep{Agarwal2013Model,Alqatawna2023Forecasting,Wong2005BoxJenkins,Wong2011ManpowerDemand}. Aggregate labor-market forecasts do not directly yield project- or task-level demand trajectories because moving across aggregation levels requires explicit disaggregation or forecast reconciliation \citep{Hyndman2011OptimalCombination}. Furthermore, construction labor demand at lower levels depends on project-specific characteristics not represented in aggregate labor-market models \citep{Wong2008Multivariate}. Consequently, the reviewed frameworks do not directly solve the variable-horizon and task-level workforce allocation problems considered in this study.

Recent studies have increasingly adopted tree-based ensembles and neural networks to overcome these limitations. \citep{Cao2024Workforce} compared several ML models for maintenance and repair projects in the Hong Kong construction industry. They reported that XGBoost performed the best in workforce demand forecasting, whereas LSTM achieved the best performance in forecasting the total construction cost. \citep{Eichenseer2025DeliveryPositions} forecast daily delivery positions to support workforce planning in logistics operations. An ensemble combining RF, a light gradient boosting machine (LGBM), XGBoost, LR, NLinear, and a time-series dense encoder (TiDE) \citep{Das2023TiDE} performed the best over short forecast horizons, whereas the LGBM achieved the highest performance over longer horizons. \citep{Ogungbire2025Workforce} compared ML regression models to forecast the demand for construction engineers and inspection personnel at a transportation infrastructure agency, and found that RF produced the best results. These studies focused primarily on predictive modeling and demonstrated an improved capacity to represent nonlinear relationships and complex patterns. Nevertheless, they retain fixed forecast horizons and do not explicitly incorporate operational constraints into their forecasting procedures.

Overall, previous studies have refined model structures to improve forecast accuracy. However, their methods are not directly applicable to the problems considered here for two reasons. First, most studies generate forecasts for one day, five business days, or another pre-specified fixed horizon. Instead, this study forecasts workforce demand until the completion of each task, resulting in different forecast horizons across tasks. Second, conventional regression models such as ARIMA and LR, as well as ensemble trees and neural networks, generally produce unconstrained point forecasts based on conditional expectations. However, they did not explicitly account for the additional constraints arising in construction operations. By contrast, this study imposes the operational requirement that the sum of the forecast sequence equals the planned cumulative workforce input.

\section{Methodology}
\label{sec:sec3}

\begin{figure}[htbp]
    \centering
    \includegraphics[width=\linewidth]{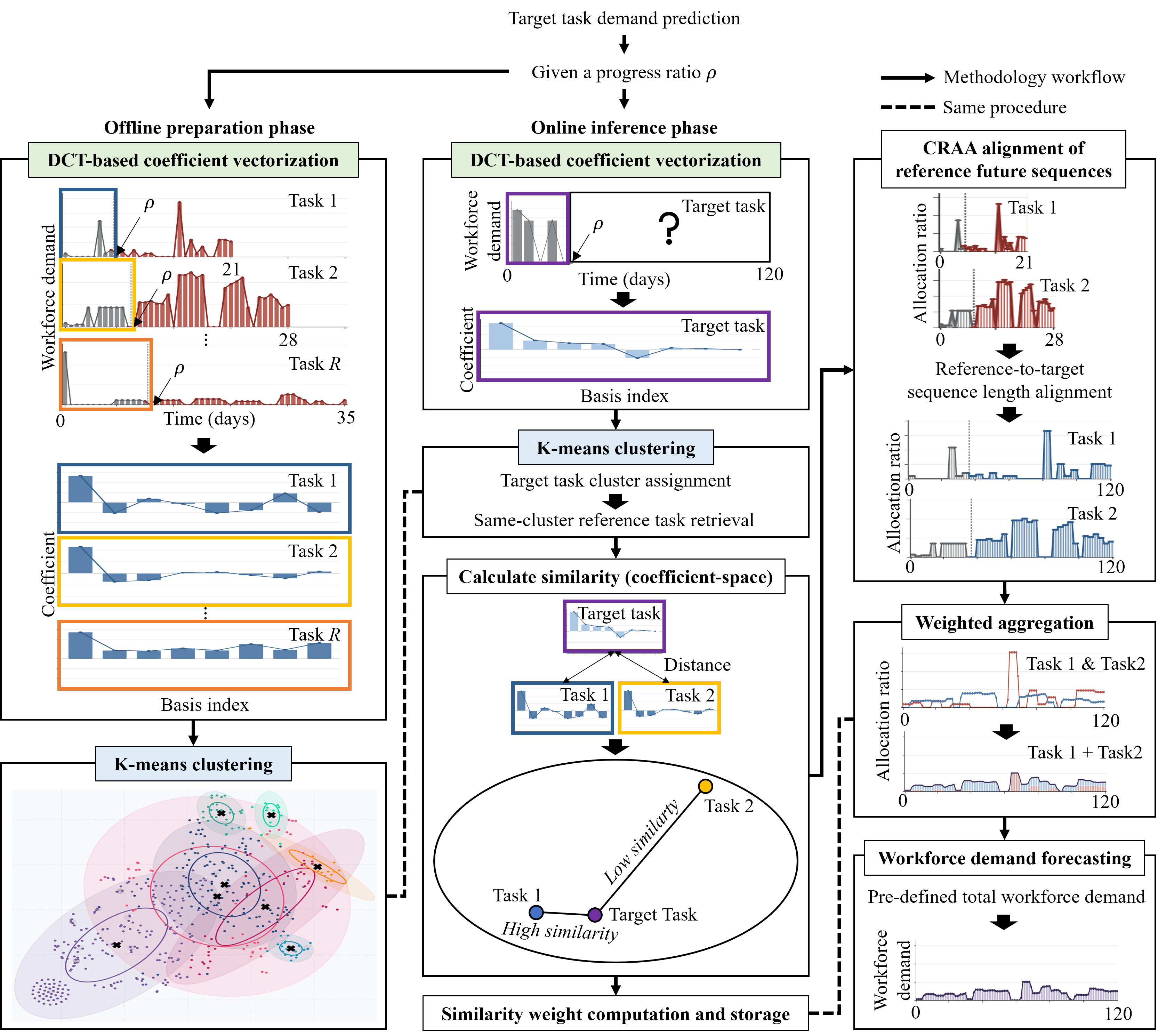}
    \caption{Framework of the proposed method. The framework comprises an offline preparation phase and online inference phase. Once offline preparation has been completed, it need not be repeated during online inference unless the data is updated.}
    \label{fig:fig2}
\end{figure}

The proposed constraint-preserving residual allocation forecasting (CP-RAF) method identifies completed reference tasks whose observed workforce demand profiles resemble that of a target task and combines their residual allocation patterns to forecast the target task’s future demand. The CP-RAF comprises an offline preparation phase and an online inference phase. During offline preparation, the method stores a discrete cosine transform (DCT)-based shape coefficient vector, cluster assignment, and normalized residual-allocation vector for each eligible reference task. During online inference, the observed demand profile of the target task is represented in the same coefficient space and candidate references are retrieved from the cluster assigned to the target. The allocation vector of each reference task was then aligned with the residual horizon of the target task using a cumulative residual allocation alignment (CRAA), and the aligned vectors were aggregated using similarity-based weights. Scaling and the resulting allocation proportions by the target task’s known residual planned amount produce a non-negative forecast whose cumulative sum equals the planned residual amount.Figure \ref{fig:fig2} presents the overall CP-RAF framework, and Table \ref{tab:tab1} defines the principal notations.

\begin{longtable}[]{@{}
		>{\raggedright\arraybackslash}p{(\linewidth - 2\tabcolsep) * \real{0.1898}}
		>{\raggedright\arraybackslash}p{(\linewidth - 2\tabcolsep) * \real{0.8102}}@{}}
	\caption{Principal notation used in CP-RAF.}\label{tab:tab1}\\
	\toprule\noalign{}
	\begin{minipage}[b]{\linewidth}\raggedright
		Notation
	\end{minipage} & \begin{minipage}[b]{\linewidth}\raggedright
		Description
	\end{minipage} \\
	\midrule\noalign{}
	\endfirsthead
	\toprule\noalign{}
	\begin{minipage}[b]{\linewidth}\raggedright
		Notation
	\end{minipage} & \begin{minipage}[b]{\linewidth}\raggedright
		Description
	\end{minipage} \\
	\midrule\noalign{}
	\endhead
	\bottomrule\noalign{}
	\endlastfoot
	\(r\) & Task index. \\
	\(i\) & Target task index. \\
	\(q\) & Reference task index. \\
	\(m\) & Daily observation index for task \(r\), \(m = 1,\ldots,T_{r}\). \\
	\(\mathbf{y}_{r}\) & Complete daily workforce demand time-series of task \(r\); \(\mathbf{y}_{r} = \left( y_{r,1},\ldots,y_{r,T_{r}} \right)^{T}\). \\
	\(y_{r,m}\) & Workforce demand recorded for task \(r\) on day \(m\). \\
	\(T_{r}\) & Planned total time-series length of task \(r\), defined as the number of days from task start to completion. \\
	\(n_{r}\) & Index of the most recent observation available for task \(r\) at the forecast origin. \\
	\(L_{r}\) & Remaining forecast horizon of task \(r\) at the forecast origin, \(L_{r} = T_{r} - n_{r}\). \\
	\(\rho\) & Task progress ratio at the forecast origin, provided as an observed input variable. \\
	\(Y_{r}\) & Known planned cumulative workforce input for task \(r\). \\
	\(M_{i}\) & Remaining planned workforce input of target task \(i\), \(M_{i} = Y_{i} - \sum_{m = 1}^{n_{i}}y_{i,m}\). \\
	\(\mathbf{\theta}_{r}\) & DCT-based shape coefficient vector representing the normalized observed workforce demand profile of task \(r\). \\
	\(\mathcal{C}_{i}\) & Set of reference tasks selected for target task \(i\). \\
	\(w_{iq}\) & Normalized similarity-based weight assigned to reference task \(q \in \mathcal{C}_{i}\) when forecasting target task \(i\), where \(\sum_{q \in \mathcal{C}_{i}}w_{iq} = 1\). \\
	\(\mathbf{a}_{q}\) & Normalized workforce allocation-proportion vector over the residual period of reference task \(q\). \\
	\({\overline{\mathbf{a}}}_{q \rightarrow i}\) & Allocation vector of reference task \(q\) aligned to the residual horizon of target task \(i\) using CRAA, whereas the horizon is \(L_{i}:\) \({\overline{\mathbf{a}}}_{q \rightarrow i} = \left( {\overline{a}}_{q \rightarrow i,1},\ldots,{\overline{a}}_{q \rightarrow i,L_{i}} \right)^{T}\). \\
	\(\mathbf{p}_{i}\) & The aligned allocation vectors of reference tasks \(q\) are combined using similarity-based weights \(w_{iq}\) to obtain the predicted residual allocation-proportion vector of target task \(i:\) \(\mathbf{p}_{i} = \left( p_{i,1},\ldots,p_{i,L_{i}} \right)^{T}\). \\
	\({\widehat{\mathbf{y}}}_{i}\) & Final workforce demand forecast vector over the residual period of target task \(i:\) \({\widehat{\mathbf{y}}}_{i} = \left( {\widehat{y}}_{i,n_{i} + 1},\ldots,{\widehat{y}}_{i,n_{i} + L_{i}} \right)^{T}\). \\
\end{longtable}

\subsection{Problem formulation}
\label{sec:sec3-1}
For task \(r\), let the daily workforce demand time series be

\begin{equation}
	\label{eq:1}
	\mathbf{y}_{r} = \left( y_{r,1},\ldots,y_{r,T_{r}} \right)^{T} \in \mathbb{R}_{+}^{T_{r}}
\end{equation}

Here, \(T_{r}\) denotes the full planned series length and \(Y_{r} > 0\) is the planned cumulative workforce input known at the forecast origin. Let \(n_{r}\) denote the index of the last observation available in the task snapshot corresponding to the recorded progress ratio \(\rho\). The residual horizon and remaining planned workforce input are defined as

\begin{equation}
	\label{eq:2}
	L_{r} = T_{r} - n_{r},\ \ M_{i} = Y_{i} - \sum_{m = 1}^{n_{i}}y_{i,m}
\end{equation}

The forecasting problem is subject to the following conditions:

\begin{equation}
	\label{eq:3}
	1 \leq n_{i} < T_{i},\ \ 0 \leq \sum_{m = 1}^{n_{i}}y_{i,m} \leq Y_{i},\ \ L_{i} \geq 1
\end{equation}

Therefore, \(M_{i} \geq 0\). CP-RAF estimates a residual allocation-proportion vector \(\mathbf{p}_{i} = \left\{ p_{i,1},\ldots,p_{i,L_{i}} \right\} \in \mathbb{R}_{+}^{L_{i}}\) whose entries sum to one. Section \ref{sec:sec3-3} defines how \(\mathbf{p}_{i}\) into the final workforce demand forecast.

\subsection{Observed-shape representation and reference retrieval}
\label{sec:sec3-2}

To compare tasks with different demand scales and observed sequence lengths, the CP-RAF normalizes the observed prefix by the planned cumulative workforce input and represents its shape using a common DCT basis. Let \(\mathbf{y}_{r,1:n_{r}} = \left( y_{r,1},\ldots,y_{r,n_{r}} \right)^{T}\) and let \(\mathbf{\Phi}_{r} \in \mathbb{R}^{n_{r} \times D_{DCT}}\) denote the DCT design matrix evaluated on the normalized time grid of the observed prefix. The shape coefficient vector is estimated as

\begin{equation}
	\label{eq:4}
	\mathbf{\theta}_{r} = \underset{\theta \in \mathbb{R}^{D_{DCT}}}{argmin}\left\{ \left\| \frac{\mathbf{y}_{r,1:n_{r}}}{Y_{r}} - \mathbf{\Phi}_{r}\mathbf{\theta} \right\|_{2}^{2} + \lambda_{ridge}\left\| \mathbf{\theta} \right\|_{2}^{2} \right\},\ \ \lambda_{ridge} > 0
\end{equation}

All the tasks used the same DCT type, scaling convention, number of coefficients, and preprocessing procedures.

During the offline preparation, K-means clustering was fitted to the coefficient vectors of the completed reference tasks. During online inference, the target coefficient vector \(\mathbf{\theta}_{i}\) is assigned to the nearest fitted centroid. The final candidate set \(\mathcal{C}_{i}\) comprises the reference tasks that satisfy the following conditions:

\begin{itemize}
	\item
	The reference task was completed and available before the target forecast origin.
	\item
	The reference task belongs to the same fitted cluster and the target task.
\end{itemize}

The cosine similarity between target task \(i\) and candidate reference task \(q \in \mathcal{C}_{i}\) is

\begin{equation}
	\label{eq:5}
	s_{iq} = \frac{{\mathbf{\theta}_{i}}^{T}\mathbf{\theta}_{q}}{\left\| \mathbf{\theta}_{i} \right\|_{2}\left\| \mathbf{\theta}_{q} \right\|_{2} + \varepsilon},\ \ q \in \mathcal{C}_{i}
\end{equation}

Given the temperature parameter \(\tau > 0\), the similarity-based weights are defined as

\begin{equation}
	\label{eq:6}
	w_{iq} = \frac{\exp\left( \frac{s_{iq}}{\tau} \right)}{\sum_{q^{'} \in \mathcal{C}_{i}}{\exp\left( \frac{s_{iq^{'}}}{\tau} \right)}},\ \ q \in \mathcal{C}_{i}
\end{equation}

Accordingly, \(w_{iq} > 0\) and \(\sum_{q \in \mathcal{C}_{i}}w_{iq} = 1\).

\subsection{CRAA-based variable-horizon forecasting}
\label{sec:sec3-3}

The reference and target tasks may have different residual horizons even at the same progress ratio. The CRAA retains the residual allocation pattern of each reference task while mapping it to the residual horizon of the target task.

For candidate reference task \(q \in \mathcal{C}_{i}\), define the normalized residual-allocation vector

\begin{equation}
	\label{eq:7}
	\mathbf{a}_{q} = \frac{\left( y_{q,n_{q} + 1},\ldots,y_{q,T_{q}} \right)^{T}}{\sum_{h = 1}^{L_{q}}y_{q,n_{q} + h}},\ \ \mathbf{1}^{T}\mathbf{a}_{q} = 1
\end{equation}

where \(\mathbf{a}_{q} \in \mathbb{R}_{+}^{L_{q}}\). Let \(A_{q}:\lbrack 0,1\rbrack \rightarrow \lbrack 0,1\rbrack\) denote the piece-wise linear cumulative allocation curve with the following grid values:

\begin{equation}
	\label{eq:8}
	A_{q}(0) = 0,\ \ A_{q}\left( \frac{h}{L_{q}} \right) = \sum_{b = 1}^{h}a_{q,b},\ \ h = 1,\ldots,L_{q}
\end{equation}

The \(j\)-th component of the reference allocation vector aligned with the target horizon is the increment in this curve over the \(j\)-th target interval.

\begin{equation}
	\label{eq:9}
	{\overline{a}}_{q \rightarrow i,j} = A_{q}\left( \frac{j}{L_{i}} \right) - A_{q}\left( \frac{j - 1}{L_{i}} \right),\ \ j = 1,\ldots,L_{i}
\end{equation}

where \({\overline{\mathbf{a}}}_{q \rightarrow i} = \left( {\overline{a}}_{q \rightarrow i,1},\ldots,{\overline{a}}_{q \rightarrow i,L_{i}} \right)^{T}\). CRAA is illustrated in Figure \ref{fig:fig3}, and its pseudocode is provided in Appendix \ref{sec:appA}.

\begin{figure}[htbp]
	\centering
	\includegraphics[width=\linewidth]{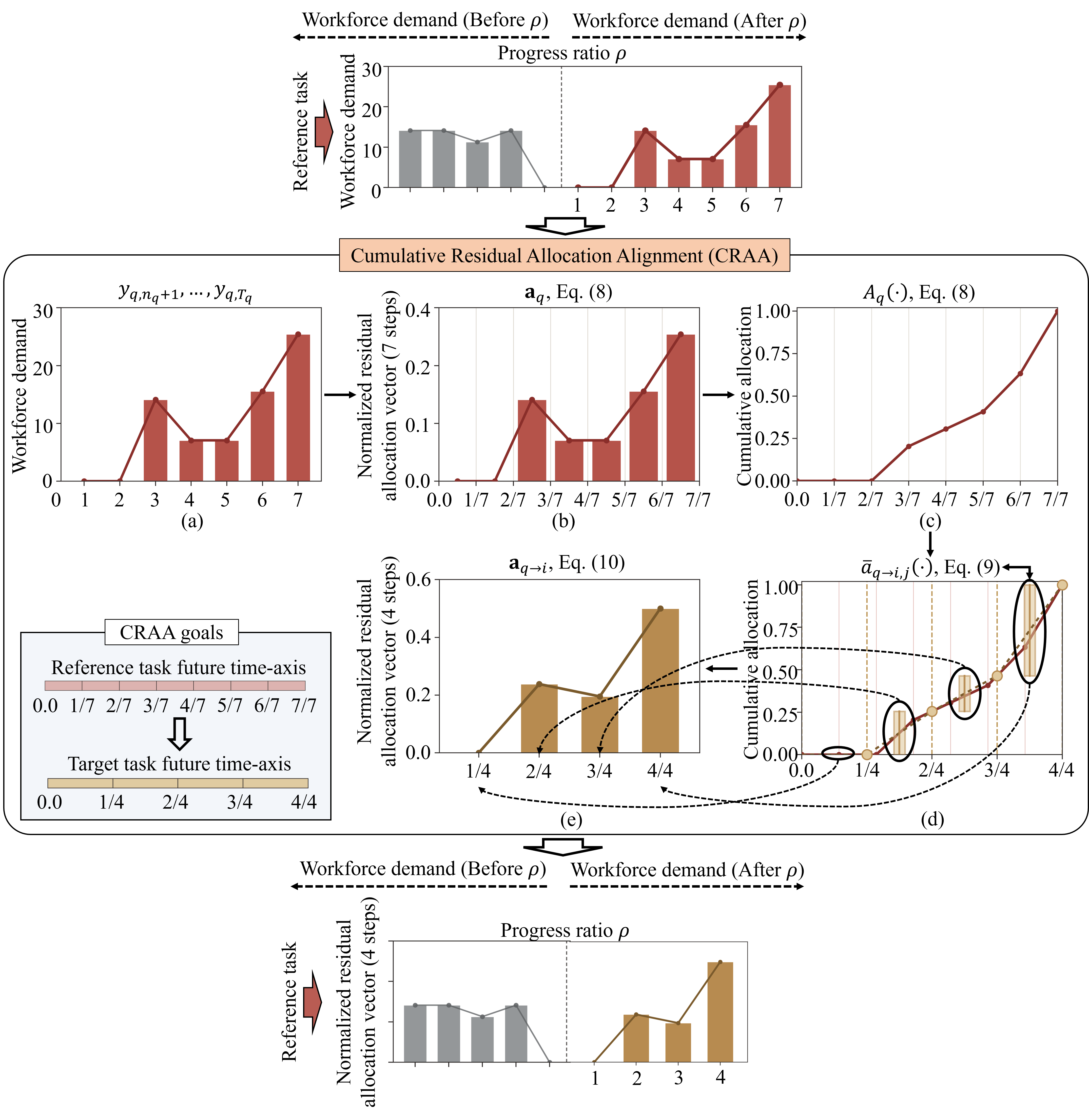}
	\caption{Example of CRAA aligning the seven-day residual allocation profile of a reference task with the four-day residual horizon of a target task.}
	\label{fig:fig3}
\end{figure}

Since \(\mathbf{a}_{q}\) is nonnegative, \(A_{q}\) is nondecreasing. Therefore,

\begin{equation}
	\label{eq:10}
	{\overline{a}}_{q \rightarrow i,j} \geq 0,\ \ \sum_{j = 1}^{L_{i}}{\overline{a}}_{q \rightarrow i,j} = A_{q}(1) - A_{q}(0) = 1
\end{equation}

The residual allocation proportion vector of the target task was computed as a similarity-weighted average.

\begin{equation}
	\label{eq:11}
	\mathbf{p}_{i} = \sum_{q \in \mathcal{C}_{i}}{w_{iq}{\overline{\mathbf{a}}}_{q \rightarrow i}}
\end{equation}

Because the weights and the aligned reference vectors lie on their respective probability simplices, \(\mathbf{p}_{i} \in \mathbb{R}_{+}^{L_{i}}\) and \(\mathbf{1}^{T}\mathbf{p}_{i} = 1\). The residual forecast is

\begin{equation}
	\label{eq:12}
	{\widehat{\mathbf{y}}}_{i} = M_{i}\mathbf{p}_{i},\ \ {\widehat{y}}_{i,n_{i} + j} = M_{i}p_{i,j},\ \ j = 1,\ldots,L_{i}
\end{equation}

Thus, CP-RAF preserves the remaining planned workforce input \(M_{i}\),

\begin{equation}
	\label{eq:13}
	\sum_{j = 1}^{L_{i}}{\widehat{y}}_{i,n_{i} + j} = M_{i}\sum_{j = 1}^{L_{i}}p_{i,j} = M_{i}
\end{equation}

as well as the full planned cumulative workforce input:

\begin{equation}
	\label{eq:14}
	\sum_{m = 1}^{n_{i}}y_{i,m} + \sum_{j = 1}^{L_{i}}{\widehat{y}}_{i,n_{i} + j} = \sum_{m = 1}^{n_{i}}y_{i,m} + M_{i} = Y_{i}
\end{equation}

Therefore, (\refeq{eq:14}) satisfies the aggregate constraint defined in (\refeq{eq:2}).

\section{Experiments}
\label{sec:sec4}

This section evaluates the CP-RAF in two settings: fixed-horizon forecasting, which permits a direct comparison with established baselines, and variable-horizon forecasting over the remaining duration of each task, which reflects the operational setting that motivates the study. The evaluation also examines the statistical significance, predicted trajectory behavior, model interpretability, sensitivity, and computational cost.

\subsection{Dataset description}
\label{sec:sec4-1}

This study used task-level daily workforce deployment data collected over 447 days, from Jun 1, 2024, to August 21, 2025, at the construction site of the S shipyard in Geoje, Republic of Korea. The dataset comprises 770 tasks. After arranging the observations for each task at daily intervals from its start date to completion date, the resulting dataset contained 93,915 task-day workforce demand observations.

The source data consisted of work-performance records manually completed by onsite personnel. During pre-processing, the observation period for each task was reconstructed as a continuous daily time series. Days without workforce deployment were assigned a value of zero. Dates absent from source records, including those resulting from omitted manual entries, were treated as days without workforce deployment and assigned a value of zero.

The prediction target was the daily workforce demand for each task, which was defined as the number of workers assigned to a given task on a given day. Because the observation period differed across tasks, the resulting time series had unequal lengths. Workforce demand levels and temporal dynamics also vary according to the task scale and characteristics. Table \ref{tab:appB1} and \ref{tab:appB2} in Appendix \ref{sec:appB} summarize the composition of the training, validation, and test sets, and the corresponding time-series characteristics. Table \ref{tab:appB3} reports the number of eligible test tasks and distribution of the residual forecast horizons at each progress ratio.

\subsection{Evaluation metric}
\label{sec:sec4-2}

The performance was assessed using five complementary metrics: the MAE, root mean squared error (RMSE), \(R^{2}\), individual-level root mean squared scaled error (iRMSSE), and asymmetric RMSE (aRMSE) \citep{Sheehan2026HorizonAware}. The MAE summarizes the average absolute deviation, whereas the RMSE is more sensitive to large pointwise errors. \(R^{2}\) measures the goodness-of-fit relative to the observed mean. The iRMSSE normalizes each task's error using its own temporal variation, thereby reducing the influence of differences in scale and sequence length. The aRMSE assigns greater weight to overestimation, reflecting the operational cost of allocating more workers than required.

Let \(y_{i,j}^{(\rho)}\) and \({\widehat{y}}_{i,j}^{(\rho)}\) denote the observed and predicted workforce demand, respectively, for inference task \(i\) at residual-horizon step \(j\). Here, \(i = 1,\ldots,I\) indexes the inference tasks, and \(j = 1,\ldots,L_{i}^{(\rho)}\) indexes the forecast steps within the residual horizon of task \(i\) at progress ratio \(\rho\). As the residual-horizon length can differ across tasks, the total number of evaluated task-day observations is

\begin{equation}
	\label{eq:15}
	N^{(\rho)} = \sum_{i = 1}^{I}L_{i}^{(\rho)}
\end{equation}

The prediction error is defined as

\begin{equation}
	\label{eq:16}
	e_{i,j}^{(\rho)} = y_{i,j}^{(\rho)} - {\widehat{y}}_{i,j}^{(\rho)}
\end{equation}

The MAE is computed as

\begin{equation}
	\label{eq:17}
	{MAE}^{(\rho)} = \frac{1}{N^{(\rho)}}\sum_{i = 1}^{I}{\sum_{j = 1}^{L_{i}^{(\rho)}}\left| e_{i,j}^{(\rho)} \right|}
\end{equation}

The RMSE is computed as

\begin{equation}
	\label{eq:18}
	{RMSE}^{(\rho)} = \sqrt{\frac{1}{N^{(\rho)}}\sum_{i = 1}^{I}{\sum_{j = 1}^{L_{i}^{(\rho)}}\left( e_{i,j}^{(\rho)} \right)^{2}}}
\end{equation}

The iRMSSE is computed as

\begin{equation}
	\label{eq:19}
	{iRMSSE}^{(\rho)} = \frac{1}{I}\sum_{i = 1}^{I}\frac{{RMSE}_{i}^{(\rho)}}{s_{i} + \epsilon}
\end{equation}

\begin{equation}
	\label{eq:20}
	s_{i} = \sqrt{\frac{1}{L_{i}^{(\rho)} - 1}\sum_{j = 1}^{L_{i}^{(\rho)}}\left( y_{i,j}^{(\rho)} - y_{i,j - 1}^{(\rho)} \right)^{2}}
\end{equation}

Here, the numerator is the task-level RMSE, the scale term is the first root mean square difference of the observed residual sequence, and epsilon is a small positive constant included in the numerical stability.

At progress ratio \(\rho\), the coefficient of determination \(R^{2,(\rho)}\) is defined as

\begin{equation}
	\label{eq:21}
	R^{2,(\rho)} = 1 - \frac{\sum_{i = 1}^{I}{\sum_{j = 1}^{L_{i}^{(\rho)}}\left( e_{i,j}^{(\rho)} \right)^{2}}}{\sum_{i = 1}^{I}{\sum_{j = 1}^{L_{i}^{(\rho)}}\left( y_{i,j}^{(\rho)} - {\overline{y}}^{(\rho)} \right)^{2}}}
\end{equation}

\begin{equation}
	\label{eq:22}
	{\overline{y}}^{(\rho)} = \frac{1}{N^{(\rho)}}\sum_{i = 1}^{I}{\sum_{j = 1}^{L_{i}^{(\rho)}}y_{i,j}^{(\rho)}}
\end{equation}

The aRMSE is computed as

\begin{equation}
	\label{eq:23}
	{aRMSE}^{(\rho)} = \sqrt{\frac{1}{N^{(\rho)}}\sum_{i = 1}^{I}{\sum_{j = 1}^{L_{i}^{(\rho)}}{\left( e_{i,j}^{(\rho)} \right)^{2}w\left( e_{i,j}^{(\rho)} \right)}}}
\end{equation}

\begin{equation}
	\label{eq:24}
	w\left( e_{i,j}^{(\rho)} \right) = \left\{ \begin{array}{r}
		2,\ \ e_{i,j}^{(\rho)} < 0 \\
		1,\ \ e_{i,j}^{(\rho)} \geq 0
	\end{array} \right.\
\end{equation}

Because error is defined as the observed demand minus the predicted demand, a negative error represents an overestimation. Therefore, (\refeq{eq:24}) assigns twice the weight to the squared errors caused by the overestimation.

\subsection{Experimental settings}
\label{sec:sec4-3}

The dataset was partitioned chronologically at the task level into training (70\%), validation (15\%), and testing (15\%) sets. This order preserves the prospective forecasting settings. To prevent leakage, all representation fitting, clustering, reference retrieval, hyperparameter selection, and normalization used only the information available before the relevant forecast origin; observations occurring after that origin were excluded from the model inputs.

Two evaluation cases are defined. Exp-case 1 used fixed horizons of 3, 5, 7, 10, and 15 days, allowing a direct comparison across the CP-RAF and the eight baselines. Exp-case 2 forecasts the complete remaining sequence at progress ratios \(\rho \in \left\{ 0.1,\ldots,0.9 \right\}\), so the output length varies by task and progress ratio. Because the baseline implementations assume a fixed output length, Exp-case 2 is used to assess CP-RAF's horizon and constraint-preserving capability of the CP-RAF, rather than as a direct head-to-head comparison. Table \ref{tab:tab2} summarizes these two cases.

\begin{longtable}[]{@{}
		>{\raggedright\arraybackslash}p{(\linewidth - 4\tabcolsep) * \real{0.1754}}
		>{\raggedright\arraybackslash}p{(\linewidth - 4\tabcolsep) * \real{0.1620}}
		>{\raggedright\arraybackslash}p{(\linewidth - 4\tabcolsep) * \real{0.6626}}@{}}
	\caption{Experimental design for fixed- and variable-horizon forecasting.}\label{tab:tab2}\\
	\toprule\noalign{}
	\begin{minipage}[b]{\linewidth}\raggedright
		Experiments type
	\end{minipage} & \begin{minipage}[b]{\linewidth}\raggedright
		Prediction horizon
	\end{minipage} & \begin{minipage}[b]{\linewidth}\raggedright
		Description
	\end{minipage} \\
	\midrule\noalign{}
	\endfirsthead
	\toprule\noalign{}
	\begin{minipage}[b]{\linewidth}\raggedright
		Experiments type
	\end{minipage} & \begin{minipage}[b]{\linewidth}\raggedright
		Prediction horizon
	\end{minipage} & \begin{minipage}[b]{\linewidth}\raggedright
		Description
	\end{minipage} \\
	\midrule\noalign{}
	\endhead
	\bottomrule\noalign{}
	\endlastfoot
	Exp-case 1 & 3, 5, 7, 10, 15 & Five fixed output lengths; direct comparison with eight baselines. \\
	Exp-case 2 & Variable & Nine variable-length forecasts at progress ratios \(\rho \in \left\{ 0.1,\ldots,0.9 \right\}\). Each task is forecast through its planned completion; only CP-RAF is reported because the baselines do not natively support task-specific output lengths. \\
\end{longtable}

Baselines were selected from the construction, logistics, and workforce-demand forecasting studies identified in the related work review, spanning statistical, linear, tree ensemble, recurrent neural network, and recent linear neural network approaches. The comparison set comprised the Box--Jenkins ARIMA \citep{Wong2005BoxJenkins}, LR \citep{Agarwal2013Model}, SARIMAX and LSTM \citep{Alqatawna2023Forecasting}, XGBoost \citep{Cao2024Workforce}, Hybrid LGBM and NLinear \citep{Eichenseer2025DeliveryPositions}, and RF \citep{Ogungbire2025Workforce}. Thus, neural approaches are represented by LSTM and NLinear, and the benchmark is domain-aligned rather than intended to exhaust all general-purpose time-series architectures. Table III lists the model-specific search spaces.

\begin{longtable}[]{@{}
		>{\raggedright\arraybackslash}p{(\linewidth - 4\tabcolsep) * \real{0.2593}}
		>{\raggedright\arraybackslash}p{(\linewidth - 4\tabcolsep) * \real{0.1262}}
		>{\raggedright\arraybackslash}p{(\linewidth - 4\tabcolsep) * \real{0.6146}}@{}}
	\caption{Hyperparameter search spaces for the baseline and proposed models.}\label{tab:tab3}\\
	\toprule\noalign{}
	\begin{minipage}[b]{\linewidth}\raggedright
		Author
	\end{minipage} & \begin{minipage}[b]{\linewidth}\raggedright
		Model
	\end{minipage} & \begin{minipage}[b]{\linewidth}\raggedright
		Hyper parameters
	\end{minipage} \\
	\midrule\noalign{}
	\endfirsthead
	\toprule\noalign{}
	\begin{minipage}[b]{\linewidth}\raggedright
		Author
	\end{minipage} & \begin{minipage}[b]{\linewidth}\raggedright
		Model
	\end{minipage} & \begin{minipage}[b]{\linewidth}\raggedright
		Hyper parameters
	\end{minipage} \\
	\midrule\noalign{}
	\endhead
	\bottomrule\noalign{}
	\endlastfoot
	\citep{Wong2005BoxJenkins} & Box--Jenkins & \(p,q \in \left\{ 0,\ldots,5 \right\}\), \(d \in \left\{ 0,1,2 \right\}\) \\
	\citep{Agarwal2013Model} & LR & - \\
	\citep{Alqatawna2023Forecasting} & SARIMAX & \((p,d,q) \times (P,D,Q)_{s}\), where \(p,q \in \left\{ 0,1,2,3 \right\}\), \(d \in \left\{ 0,1,2 \right\}\), \(P,Q \in \left\{ 0,1,2 \right\}\), \(D \in \left\{ 0,1 \right\}\), \(s = 7\) \\
	\citep{Alqatawna2023Forecasting} & LSTM & \(h \in \left\{ 32,64,128,256 \right\},\) \(L \in \left\{ 1,2 \right\}\), \(r_{drop} \in \left\{ 0.0,0.1,0.2,0.3 \right\}\), \(\eta \in \left\{ 10^{- 4},10^{- 3} \right\}\), \(\lambda_{reg} \in \left\{ 0,10^{- 6},10^{- 5},10^{- 4} \right\}\), \(B \in \left\{ 32,64,128 \right\}\) \\
	\citep{Cao2024Workforce} & XGBoost & \(N_{est} \in \left\{ 200,300,500,700,1000 \right\}\), \(d_{\max} \in \left\{ 2,3,4,5,6,8 \right\}\), \(p_{col}\sim Unif(0.6,1.0)\), \(w_{\min}^{child} \in \left\{ 1,3,5,10 \right\}\), \(\gamma \in \left\{ 0,0.01,0.1,1.0 \right\}\), \(\alpha \in \left\{ 0,10^{- 4},10^{- 3},10^{- 2},0.1 \right\}\), \(\lambda \in \left\{ 0,10^{- 4},10^{- 3},10^{- 2},0.1 \right\}\) \\
	\citep{Eichenseer2025DeliveryPositions} & Hybrid LGBM & \(N_{est} \in \left\{ 200,300,500,700,1000 \right\}\), \(N_{leaf} \in \left\{ 15,31,63,127 \right\}\), \(d_{\max} \in \left\{ - 1,5,10,15 \right\}\), \(n_{\min}^{child} \in \left\{ 10,20,40,80 \right\}\), \(p_{col}\sim Unif(0.6,1.0)\), \(\alpha \in \left\{ 0,10^{- 4},10^{- 3},10^{- 2},0.1 \right\}\), \(\lambda \in \left\{ 0,10^{- 4},10^{- 3},10^{- 2},0.1 \right\}\) \\
	\citep{Eichenseer2025DeliveryPositions} & NLinear & \(\eta \in \left\{ 10^{- 4},5 \times 10^{- 3},10^{- 3} \right\}\), \(\lambda_{reg} \in \left\{ 0,10^{- 6},10^{- 5},10^{- 4} \right\}\), \(B \in \left\{ 64,128,256 \right\}\) \\
	\citep{Ogungbire2025Workforce} & RF & \(N_{est} \in \left\{ 200,300,500,700,1000 \right\}\), \(d_{\max} \in \left\{ 5,8,10,15,20 \right\}\), \(n_{\min}^{split} \in \left\{ 2,5,10 \right\}\), \(n_{\min}^{leaf} \in \left\{ 1,2,4 \right\}\), \(p_{col} \in \left\{ 0.5,0.8 \right\}\) \\
	& Proposed & \(k \in \left\{ 5,6,7,8,9,10,11 \right\}\), \(d_{dct} \in \left\{ 4,6,8,12,16,24,32 \right\}\) \\
\end{longtable}

\ref{tab:tab3} presents the standard parameterization for each implementation. For the Box--Jenkins model, \(p,d,q\) denote nonseasonal orders; specifically, \(p,d\), and \(q\) are autoregressive, differencing, and moving average orders, respectively. In SARIMAX, \(P,D,Q,s\) denote the seasonal autoregressive, differencing, moving average, and periodicity terms, respectively\footnote{\url{https://www.statsmodels.org/stable/generated/statsmodels.tsa.arima.model.ARIMA.html}}. For LSTM, \(h\) is the hidden state dimension, \(L\) the number of recurrent layers, and \(r_{drop}\) the dropout rate\footnote{\url{https://docs.pytorch.org/docs/2.12/generated/torch.nn.LSTM.html}}; \(\eta\) and \(\lambda_{reg}\) are the learning rate and weight decay coefficient\footnote{\url{https://docs.pytorch.org/docs/2.12/generated/torch.optim.Adam.html}}, and \(B\) is the mini-batch size. For XGBoost and LGBM, \(N_{est},d_{\max},p_{col},\alpha,\lambda\) denote the number of trees, maximum depth, column-subsampling ratio, and L1 and L2 regularization coefficients. In XGBoost, \(w_{\min}^{child}\) is the minimum child weight and \(\gamma\) is the minimum loss reduction required for a split\footnote{\url{https://xgboost.readthedocs.io/en/latest/python/python_api.html}}. In the LGBM, \(N_{leaf}\) and \(n_{\min}^{child}\) are the maximum number of leaves and the minimum number of observations in a child or leaf, respectively\footnote{\url{https://lightgbm.readthedocs.io/en/latest/pythonapi/lightgbm.LGBMRegressor.html}}. For RF, \(N_{est},d_{\max},n_{\min}^{split},n_{\min}^{leaf}\) denote the number of trees, maximum depth, minimum samples required to split an internal node, and minimum samples in a leaf, respectively; \(p_{col}\) is the proportion of candidate features considered at each split\footnote{\url{https://scikit-learn.org/stable/modules/generated/sklearn.ensemble.RandomForestRegressor.html}}. NLinear\footnote{\url{https://github.com/cure-lab/LTSF-Linear}} uses \(\eta,\lambda_{reg},B\) for learning rate, weight decay, and batch size. Finally, CP-RAF uses \(k\) as the number of clusters and \(d_{dct}\) as the DCT coefficient dimension. LR is an unregularized ordinary least-squares baseline, and therefore does not tune the hyperparameters. NLinear was implemented using the code released by the authors.

For each model and evaluation setting, 50 hyperparameter configurations were sampled using random search. The configuration with the lowest validation MAE is selected and fixed for testing. Selected configurations are presented in Appendix \ref{sec:appC}.

\subsection{Experimental results}
\label{sec:sec4-4}

The results were organized around six questions: 1) how CP-RAF compares with the eight baselines under fixed horizons, 2) whether the observed improvements are statistically significant, 3) how the predicted trajectories differ qualitatively, 4) whether CP-RAF remains accurate when each task has its own residual horizon, 5) what its clusters and sensitivity patterns reveal about the model, and 6) whether its inference latency is compatible with operational use.

\begin{longtable}[]{@{}
		>{\raggedright\arraybackslash}p{(\linewidth - 12\tabcolsep) * \real{0.2461}}
		>{\raggedright\arraybackslash}p{(\linewidth - 12\tabcolsep) * \real{0.1622}}
		>{\raggedright\arraybackslash}p{(\linewidth - 12\tabcolsep) * \real{0.1108}}
		>{\raggedright\arraybackslash}p{(\linewidth - 12\tabcolsep) * \real{0.1252}}
		>{\raggedright\arraybackslash}p{(\linewidth - 12\tabcolsep) * \real{0.1342}}
		>{\raggedright\arraybackslash}p{(\linewidth - 12\tabcolsep) * \real{0.0964}}
		>{\raggedright\arraybackslash}p{(\linewidth - 12\tabcolsep) * \real{0.1252}}@{}}
	\caption{Fixed-horizon forecasting performance in Exp-case 1.}\label{tab:tab4}\\
	\toprule\noalign{}
	\begin{minipage}[b]{\linewidth}\raggedright
		Prediction horizon
	\end{minipage} & \begin{minipage}[b]{\linewidth}\raggedright
		Model
	\end{minipage} & \begin{minipage}[b]{\linewidth}\raggedright
		MAE
	\end{minipage} & \begin{minipage}[b]{\linewidth}\raggedright
		RMSE
	\end{minipage} & \begin{minipage}[b]{\linewidth}\raggedright
		iRMSSE
	\end{minipage} & \begin{minipage}[b]{\linewidth}\raggedright
		R\textsuperscript{2}
	\end{minipage} & \begin{minipage}[b]{\linewidth}\raggedright
		aRMSE
	\end{minipage} \\
	\midrule\noalign{}
	\endfirsthead
	\toprule\noalign{}
	\begin{minipage}[b]{\linewidth}\raggedright
		Prediction horizon
	\end{minipage} & \begin{minipage}[b]{\linewidth}\raggedright
		Model
	\end{minipage} & \begin{minipage}[b]{\linewidth}\raggedright
		MAE
	\end{minipage} & \begin{minipage}[b]{\linewidth}\raggedright
		RMSE
	\end{minipage} & \begin{minipage}[b]{\linewidth}\raggedright
		iRMSSE
	\end{minipage} & \begin{minipage}[b]{\linewidth}\raggedright
		R\textsuperscript{2}
	\end{minipage} & \begin{minipage}[b]{\linewidth}\raggedright
		aRMSE
	\end{minipage} \\
	\midrule\noalign{}
	\endhead
	\bottomrule\noalign{}
	\endlastfoot
	3 & Box--Jenkins & 45.251 & 323.008 & \underline{0.834} & 0.024 & 423.560 \\
	& LR & 55.446 & 315.431 & 8.665 & 0.004 & 445.201 \\
	& SARIMAX & 47.856 & 328.649 & 0.871 & 0.080 & 442.179 \\
	& LSTM & 43.774 & 303.539 & 2.014 & 0.126 & 405.164 \\
	& XGBoost & 41.681 & 292.506 & 1.481 & 0.188 & 399.381 \\
	& Hybrid LGBM & 41.248 & 288.605 & 1.703 & \underline{0.209} & 396.816 \\
	& NLinear & 44.183 & \underline{289.177} & 2.270 & 0.207 & \underline{382.646} \\
	& \textbf{RF} & \textbf{40.449} & \textbf{286.939} & 1.226 & \textbf{0.218} & 396.519 \\
	& Proposed & \underline{40.590} & 289.918 & \textbf{0.739} & 0.166 & \textbf{365.815} \\ \midrule\noalign{}
	5 & Box--Jenkins & 45.880 & 324.073 & \underline{0.845} & 0.021 & 425.342 \\
	& LR & 55.503 & 315.509 & 8.678 & 0.004 & 445.310 \\
	& SARIMAX & 48.911 & 337.494 & 1.022 & 0.006 & 452.655 \\
	& LSTM & 41.342 & 294.030 & 2.152 & 0.180 & 397.848 \\
	& XGBoost & \underline{40.906} & \underline{284.350} & 1.546 & \underline{0.233} & 392.177 \\
	& Hybrid LGBM & 41.973 & 287.846 & 1.839 & 0.214 & 397.270 \\
	& NLinear & 45.858 & 293.881 & 2.173 & 0.181 & \underline{387.093} \\
	& RF & 41.473 & 287.580 & 1.253 & 0.215 & 396.512 \\
	& \textbf{Proposed} & \textbf{39.490} & \textbf{254.403} & \textbf{0.720} & \textbf{0.359} & \textbf{331.702} \\ \midrule\noalign{}
	7 & Box--Jenkins & 62.204 & 646.184 & \underline{0.885} & -2.860 & 692.644 \\
	& LR & 55.775 & 317.176 & 8.586 & 0.004 & 447.684 \\
	& SARIMAX & 48.427 & 324.211 & 1.111 & 0.090 & 421.716 \\
	& LSTM & 45.840 & 315.233 & 2.064 & 0.066 & 426.338 \\
	& XGBoost & 44.349 & 309.331 & 1.838 & 0.100 & 425.115 \\
	& Hybrid LGBM & 44.712 & 307.462 & 1.756 & 0.114 & 424.162 \\
	& NLinear & 46.034 & 310.988 & 2.327 & 0.091 & \underline{412.163} \\
	& RF & \underline{43.903} & \underline{304.257} & 1.364 & \underline{0.129} & 418.124 \\
	& \textbf{Proposed} & \textbf{40.540} & \textbf{273.022} & \textbf{0.704} & \textbf{0.267} & \textbf{356.971} \\ \midrule\noalign{}
	10 & Box--Jenkins & 46.013 & 362.344 & \underline{0.835} & \underline{0.158} & 505.368 \\
	& LR & 57.741 & 380.353 & 8.587 & 0.003 & 537.178 \\
	& SARIMAX & 54.815 & \underline{363.225} & 1.091 & 0.172 & \textbf{464.162} \\
	& LSTM & 55.658 & 373.796 & 5.792 & 0.091 & 518.348 \\
	& XGBoost & 46.101 & 369.956 & 2.007 & 0.109 & 514.786 \\
	& Hybrid LGBM & 45.886 & 372.633 & 1.800 & 0.096 & 519.842 \\
	& NLinear & 48.769 & 384.927 & 1.985 & 0.036 & 519.912 \\
	& RF & \underline{45.261} & 368.460 & 1.640 & 0.116 & 514.193 \\
	& \textbf{Proposed} & \textbf{42.180} & \textbf{340.492} & \textbf{0.705} & \textbf{0.206} & \underline{467.437} \\ \midrule\noalign{}
	15 & Box--Jenkins & \underline{47.179} & \underline{375.657} & \underline{0.968} & \underline{0.154} & 524.666 \\
	& LR & 59.831 & 395.796 & 8.696 & 0.003 & 559.036 \\
	& SARIMAX & 59.564 & 365.162 & 1.276 & 0.185 & \underline{463.549} \\
	& LSTM & 49.892 & 379.596 & 2.943 & 0.129 & 531.916 \\
	& XGBoost & 48.891 & 386.959 & 2.133 & 0.095 & 538.810 \\
	& Hybrid LGBM & 49.875 & 382.711 & 2.736 & 0.115 & 534.796 \\
	& NLinear & 51.030 & 377.366 & 2.355 & 0.140 & 513.229 \\
	& RF & 48.305 & 378.574 & 1.926 & 0.134 & 529.262 \\
	& \textbf{Proposed} & \textbf{41.991} & \textbf{334.704} & \textbf{0.778} & \textbf{0.293} & \textbf{448.203} \\
\end{longtable}

Table \ref{tab:tab4} reports fixed-horizon performance. Baseline rankings vary across horizons; no single baseline dominates all settings. At a 3-day horizon, RF achieves the lowest MAE (40.449), lowest RMSE (286.939), and highest \(R^{2}\) (0.218). CP-RAF is close on MAE (40.590) and RMSE (289.918) but has a lower \(R^{2}\) (0.166); however, it obtains the best iRMSSE (0.739) and aRMSE (365.815). This pattern indicates that, for the shortest horizon, RF is strongest on pooled error measures while CP-RAF performs better after task-level scaling and under the asymmetric penalty.

From days 5 to 15, CP-RAF ranked first on all five metrics. Relative to the best baseline MAE at each horizon, the reductions were 1.416, 3.363, 3.081, 5.188 on 5, 7, 10, and 15 days, respectively. Therefore, an advantage appears after the shortest horizon and persists as the forecasting task lengthens. These results support the intended use of CP-RAF for medium- and long-term workforce allocation, while showing that RF remains competitive in the shortest horizon.

To assess the statistical significance of the MAE differences in Table \ref{tab:tab4}, paired DM tests were conducted using the absolute error loss. Here, \(i = 1,\ldots,I\) indexes inference tasks, and \(j = 1,\ldots,T_{pred}\) indexes forecast steps within the fixed residual horizon of task \(i\). For model \(m\), the task-level horizon-averaged absolute error loss is

\begin{equation}
	\label{eq:25}
	\ell_{i,m} = \frac{1}{T_{pred}}\sum_{j = 1}^{T_{pred}}\left| y_{i,t_{s} + j} - {\widehat{y}}_{i,t_{s} + j}^{m} \right|
\end{equation}

Here, \(m\) identifies the forecasting model, \(t_{s}\) is the forecast origin, \({\widehat{y}}_{i,t_{s} + j}^{m}\) is the prediction for task \(i\) produced by model \(m\) at time \(t_{s} + j\), and \(y_{i,t_{s} + j}\) are the corresponding observed values.

The paired loss differential is

\begin{equation}
	\label{eq:26}
	d_i = \ell_{i,prop} - \ell_{i,base}
\end{equation}

Thus, \(d_i\) is the difference between the loss of the proposed model \(\ell_{i,prop}\) and the baseline loss \(\ell_{i,base}\). The null hypothesis was \(H_0:\mathbb{E}[d_i]=0\), indicating an equal expected absolute error loss. A negative DM statistic favors CP-RAF.

\begin{longtable}[]{@{}
		>{\raggedright\arraybackslash}p{(\linewidth - 8\tabcolsep) * \real{0.2749}}
		>{\raggedright\arraybackslash}p{(\linewidth - 8\tabcolsep) * \real{0.1814}}
		>{\raggedright\arraybackslash}p{(\linewidth - 8\tabcolsep) * \real{0.1814}}
		>{\raggedright\arraybackslash}p{(\linewidth - 8\tabcolsep) * \real{0.1814}}
		>{\raggedright\arraybackslash}p{(\linewidth - 8\tabcolsep) * \real{0.1810}}@{}}
	\caption{Diebold--Mariano tests comparing CP-RAF with the eight baselines in Exp-case 1.}\label{tab:tab5}\\
	\toprule\noalign{}
	\begin{minipage}[b]{\linewidth}\raggedright
		Prediction horizon
	\end{minipage} & \begin{minipage}[b]{\linewidth}\raggedright
		Model
	\end{minipage} & \begin{minipage}[b]{\linewidth}\raggedright
		DM statistics
	\end{minipage} & \begin{minipage}[b]{\linewidth}\raggedright
		p-value
	\end{minipage} & \begin{minipage}[b]{\linewidth}\raggedright
		Significance
	\end{minipage} \\
	\midrule\noalign{}
	\endfirsthead
	\toprule\noalign{}
	\begin{minipage}[b]{\linewidth}\raggedright
		Prediction horizon
	\end{minipage} & \begin{minipage}[b]{\linewidth}\raggedright
		Model
	\end{minipage} & \begin{minipage}[b]{\linewidth}\raggedright
		DM statistics
	\end{minipage} & \begin{minipage}[b]{\linewidth}\raggedright
		p-value
	\end{minipage} & \begin{minipage}[b]{\linewidth}\raggedright
		Significance
	\end{minipage} \\
	\midrule\noalign{}
	\endhead
	\bottomrule\noalign{}
	\endlastfoot
	3 & Box--Jenkins & -1.954 & 0.0507 & ns \\
	& LR & -6.994 & \textless{}\(1 \times 10^{- 9}\) & *** \\
	& SARIMAX & -3.308 & \textless{}\(1 \times 10^{- 9}\) & *** \\
	& LSTM & -1.555 & 0.1199 & ns \\
	& XGBoost & -0.484 & 0.6282 & ns \\
	& Hybrid LGBM & -0.230 & 0.8181 & ns \\
	& NLinear & -1.924 & 0.0544 & ns \\
	& RF & 0.213 & 0.8317 & ns \\ \midrule\noalign{}
	5 & Box--Jenkins & -3.373 & 7.4e-04 & *** \\
	& LR & -8.823 & \textless{}\(1 \times 10^{- 9}\) & *** \\
	& SARIMAX & -5.166 & 2.4e-07 & *** \\
	& LSTM & -0.925 & 0.3552 & ns \\
	& XGBoost & -0.732 & 0.4642 & ns \\
	& Hybrid LGBM & -1.408 & 0.1591 & ns \\
	& NLinear & -3.679 & 2.3e-04 & *** \\
	& RF & -1.117 & 0.2638 & ns \\ \midrule\noalign{}
	7 & Box--Jenkins & -4.155 & 3.2e-05 & *** \\
	& LR & -8.282 & \textless{}\(1 \times 10^{- 9}\) & *** \\
	& SARIMAX & -4.276 & 1.9e-05 & *** \\
	& LSTM & -3.204 & 0.0014 & *** \\
	& XGBoost & -2.469 & 0.0136 & * \\
	& Hybrid LGBM & -2.667 & 0.0077 & ** \\
	& NLinear & -3.358 & 7.8e-04 & *** \\
	& RF & -2.178 & 0.0294 & * \\ \midrule\noalign{}
	10 & Box--Jenkins & -2.821 & 0.0048 & *** \\
	& LR & -8.930 & \textless{}\(1 \times 10^{- 9}\) & *** \\
	& SARIMAX & -6.214 & \textless{}\(1 \times 10^{- 9}\) & *** \\
	& LSTM & -8.291 & \textless{}\(1 \times 10^{- 9}\) & *** \\
	& XGBoost & -2.537 & 0.0112 & * \\
	& Hybrid LGBM & -2.471 & 0.0135 & * \\
	& NLinear & -3.890 & 1.0e-04 & *** \\
	& RF & -2.097 & 0.0360 & * \\ \midrule\noalign{}
	15 & Box--Jenkins & -4.085 & 4.4e-05 & *** \\
	& LR & -9.787 & \textless{}\(1 \times 10^{- 9}\) & *** \\
	& SARIMAX & -8.863 & \textless{}\(1 \times 10^{- 9}\) & *** \\
	& LSTM & -5.420 & 6.0e-08 & *** \\
	& XGBoost & -4.631 & 3.6e-06 & *** \\
	& Hybrid LGBM & -5.401 & 6.7e-08 & *** \\
	& NLinear & -5.739 & \textless{}\(1 \times 10^{- 9}\) & *** \\
	& RF & -4.531 & 5.9e-06 & *** \\
\end{longtable}
Note. Negative DM statistics favor CP-RAF. * p \textless{} .05, ** p \textless{} .01, *** p \textless{} .005; ns indicates no statistically significant difference.

Table \ref{tab:tab5} summarizes the results of the DM tests. On day 3, CP-RAF significantly outperformed LR and SARIMAX, whereas the differences from Box--Jenkins, LSTM, NLinear, XGBoost, Hybrid LGBM, and RF were not significant. On day 5, significant improvements were observed over the Box--Jenkins, LR, SARIMAX, and NLinear models but not over the LSTM or tree ensembles. On days 7 and 10, all eight comparisons favored CP-RAF at the 5\% level, and on day 15, all eight comparisons were significant at p \textless{} .005. Across the 40 model--horizon comparisons, 30 were considered statistically significant. Therefore, the evidence strengthens with horizon length, which is consistent with the point estimates in Table IV, although the tests establish differences in the expected absolute error loss rather than operational causality.

\begin{figure}[htbp]
	\centering
	\includegraphics[width=\linewidth]{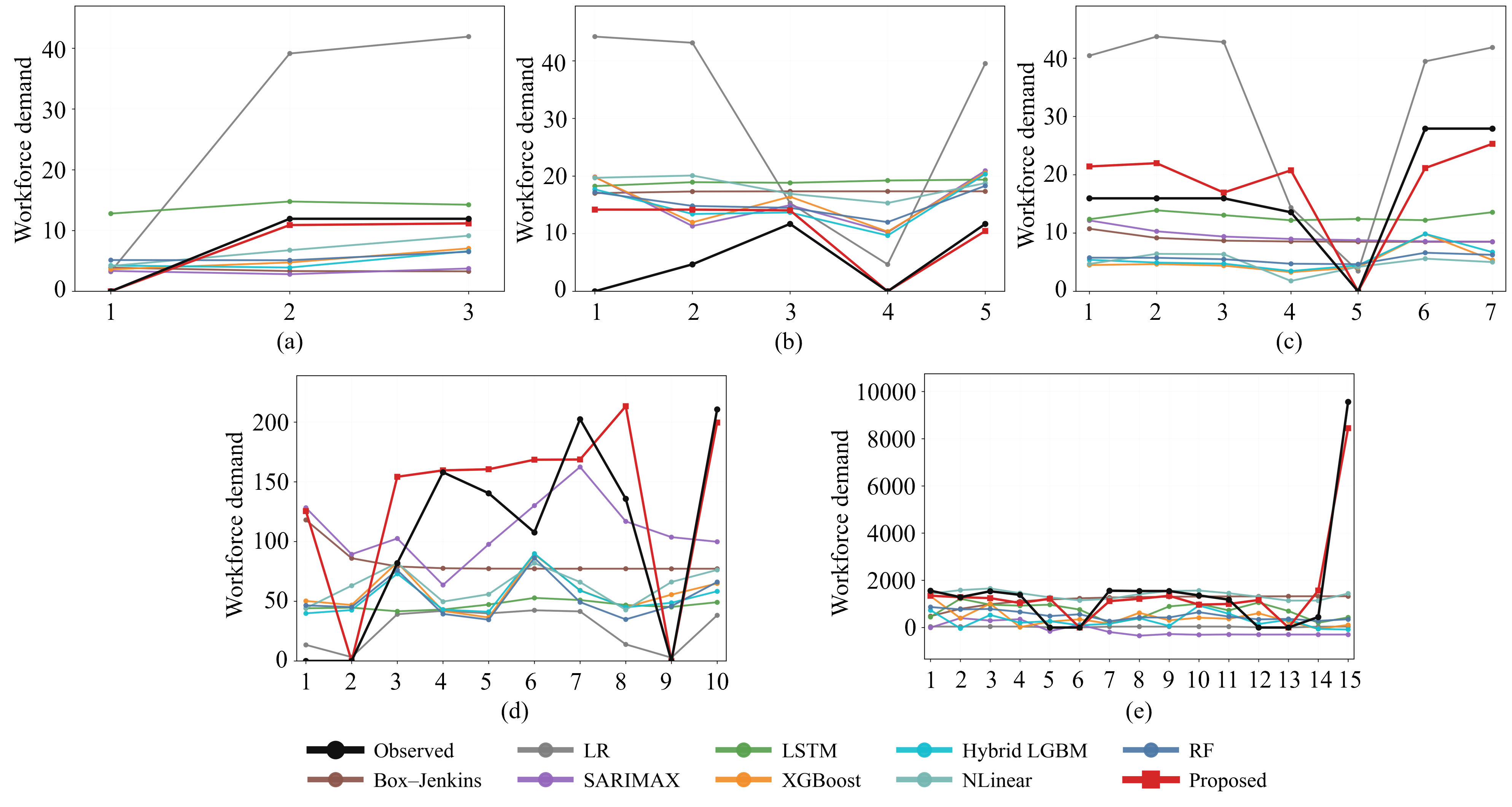}
	\caption{Visual comparison of observed and predicted trajectories for the nine models across the five fixed horizons in Exp-case 1.}
	\label{fig:fig4}
\end{figure}

Figure \ref{fig:fig4} compares the predicted trajectories at horizons of 3, 5, 7, 10, and 15 days in panels (a)--(e). Each panel shows a randomly sampled task. Therefore, the figure is illustrative rather than a substitute for the aggregate results in Table \ref{tab:tab4}, and its purpose is to show how the models respond to task-level changes in demand shape.

Across the examples, several baselines produced relatively smooth trajectories near the conditional mean and consequently underreacted to abrupt changes. Examples include the decline at step 5 in panel (c); the declines at steps 1, 2, and 9 in panel (d); and the increases at step 10 in panel (d) and step 15 in panel (e). CP-RAF follows several of these local increases and falls more closely, particularly in panels (d) and (e).

This behavior is consistent with the model design: the CP-RAF retrieves completed tasks with similar observed prefixes and transfers their residual allocation patterns instead of directly regressing each future demand level. These examples suggest that the mechanism can retain irregular and intermittent allocation shapes, including late concentrations and zero-demand intervals. However, because the panels represent individual cases, this interpretation should be read together with the quantitative results rather than generalized from the figure alone.

\begin{longtable}[]{@{}
		>{\raggedright\arraybackslash}p{(\linewidth - 14\tabcolsep) * \real{0.2296}}
		>{\raggedright\arraybackslash}p{(\linewidth - 14\tabcolsep) * \real{0.1290}}
		>{\raggedright\arraybackslash}p{(\linewidth - 14\tabcolsep) * \real{0.0900}}
		>{\raggedright\arraybackslash}p{(\linewidth - 14\tabcolsep) * \real{0.1234}}
		>{\raggedright\arraybackslash}p{(\linewidth - 14\tabcolsep) * \real{0.1118}}
		>{\raggedright\arraybackslash}p{(\linewidth - 14\tabcolsep) * \real{0.1198}}
		>{\raggedright\arraybackslash}p{(\linewidth - 14\tabcolsep) * \real{0.0852}}
		>{\raggedright\arraybackslash}p{(\linewidth - 14\tabcolsep) * \real{0.1113}}@{}}
	\caption{Variable-horizon performance of CP-RAF by progress ratio, with the median and maximum residual forecast horizons. Full horizon distributions are reported in Appendix \ref{sec:appB}.}\label{tab:tab6}\\
	\toprule\noalign{}
	\begin{minipage}[b]{\linewidth}\raggedright
		Progress ratio \(\rho\)
	\end{minipage}
	& \multicolumn{2}{l}{Prediction horizon}
	& \multicolumn{5}{l@{}}{Performance} \\
	& Median & Max & MAE & RMSE & iRMSSE
	& R\textsuperscript{2} & aRMSE \\
	\midrule\noalign{}
	\endfirsthead
	
	\toprule\noalign{}
	\begin{minipage}[b]{\linewidth}\raggedright
		Progress ratio \(\rho\)
	\end{minipage}
	& \multicolumn{2}{l}{Prediction horizon}
	& \multicolumn{5}{l@{}}{Performance} \\
	& Median & Max & MAE & RMSE & iRMSSE
	& R\textsuperscript{2} & aRMSE \\
	\midrule\noalign{}
	\endhead
	\bottomrule\noalign{}
	\endlastfoot
	0.1 & 64 & 291 & 32.877 & 313.104 & 0.795 & 0.243 & 420.794 \\
	0.2 & 57 & 259 & 32.823 & 321.144 & 0.792 & 0.248 & 432.058 \\
	0.3 & 51 & 227 & 33.173 & 326.694 & 0.788 & 0.260 & 439.919 \\
	0.4 & 44 & 194 & 33.404 & 341.165 & 0.779 & 0.245 & 458.309 \\
	0.5 & 38 & 162 & 34.383 & 356.587 & 0.783 & 0.226 & 478.401 \\
	0.6 & 31 & 130 & 33.597 & 358.875 & 0.768 & 0.272 & 486.128 \\
	0.7 & 24 & 97 & 34.001 & 404.248 & 0.748 & 0.164 & 542.202 \\
	0.8 & 18 & 65 & 35.156 & 437.848 & 0.728 & 0.210 & 591.559 \\
	0.9 & 11 & 33 & 32.741 & 468.723 & 0.686 & 0.202 & 663.260 \\
\end{longtable}

Table \ref{tab:tab6} reports CP-RAF performance at each progress ratio \(\rho\). This experiment is not a direct baseline comparison; it evaluates whether CP-RAF can forecast task-specific sequences through planned completion. At \(\rho = 0.1\), the median residual horizon is 64 days, and the maximum is 291 days. At \(\rho = 0.9\), these values decline to 11 and 33 days, respectively, as expected when more of the task has already been observed.

The MAE remained within a narrow range of 32.741--35.156 across all the progress ratios. At \(\rho = 0.1\), CP-RAF yields an MAE of 32.877 despite residual horizons of up to 291 days; corresponding MAEs at \(\rho = 0.2\) and 0.3 are 32.823 and 33.173, and the MAE at \(\rho = 0.9\) is 32.741. These values indicate a stable average absolute error across substantially different residual horizon distributions. In contrast, the RMSE and aRMSE tend to increase as the residual forecast horizon shortens. RMSE and aRMSE increase from 313.104 and 420.794 at \(\rho = 0.1\) to 468.723 and 663.260 at \(\rho = 0.9\), respectively. One possible explanation is that the CP-RAF preserves abrupt variations in workforce demand more strongly than the smoothed profiles produced by the conditional-mean baselines. When the predicted timing of a sharp demand peak differs from its actual occurrence, large pointwise errors can arise at both observed and predicted peak locations. Because the RMSE and aRMSE assign disproportionately greater weights to such large errors, these temporal mismatches can increase the two metrics, even when the overall MAE remains stable.

\begin{figure}[htbp]
	\centering
	\includegraphics[width=\textwidth]{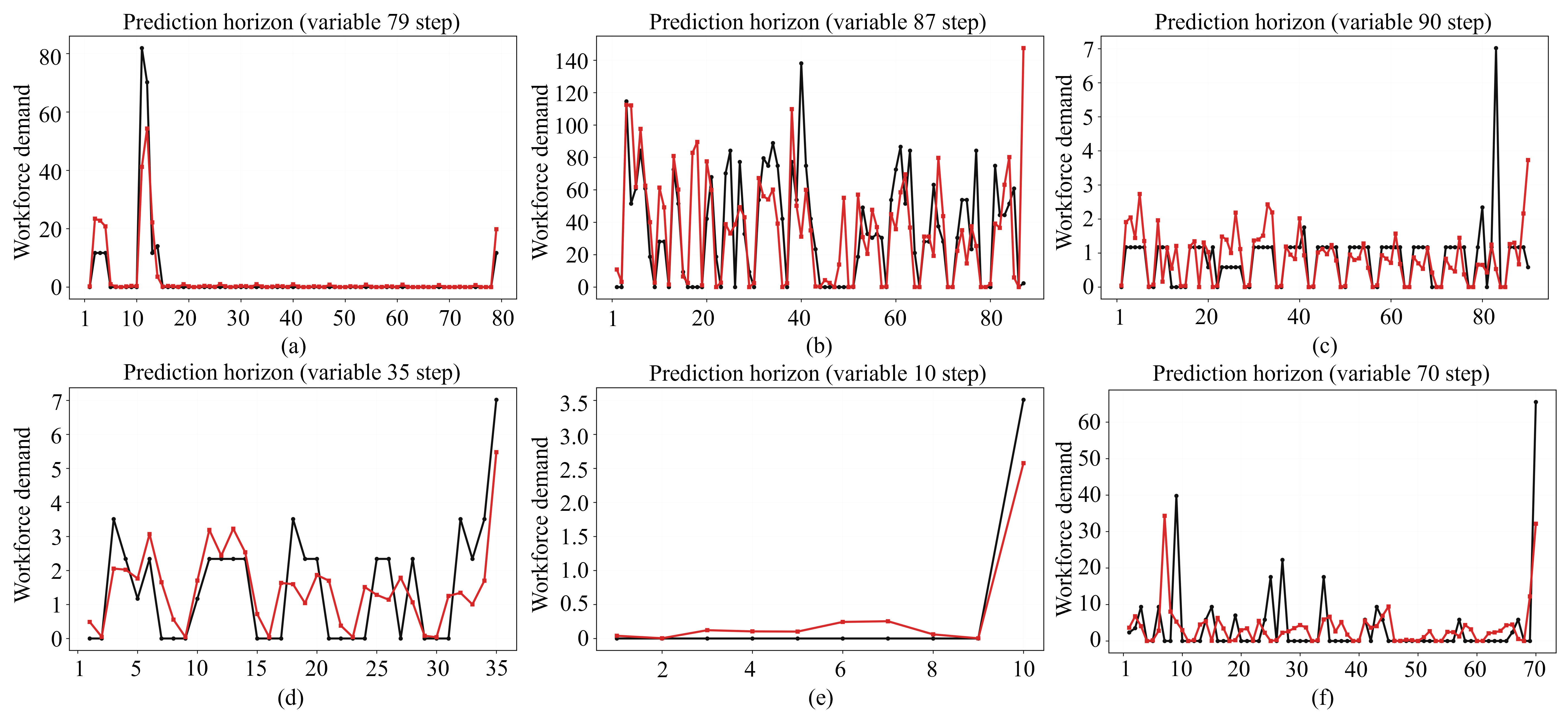}
	\caption{Illustrative variable-horizon forecasts produced by CP-RAF in Exp-case 2.}
	\label{fig:fig5}
\end{figure}

Figure \ref{fig:fig5} illustrates the CP-RAF forecasts for Experiment Case 2. The six panels span the residual horizons from 10 to 90 days and demonstrate that the same procedure produces forecasts without requiring a fixed output length. Examples include several demand shapes: panel (a) is front-loaded; panels (b) and (d) distribute demand more broadly; and panels (c), (e), and (f) contain late concentrations or late spikes. The CP-RAF reproduces a broad allocation pattern in each case, including intermittent zero-demand periods and pronounced shifts in workforce inputs.

The principal failure mode is the temporal displacement of sharp events. For example, at approximately step 80 in panel (c) and step 10 in panel (f), the observed and predicted spikes occurred at different times. Even when the overall shape and cumulative allocation are similar, such a displacement can produce large pointwise errors. This observation is consistent with the increase in RMSE and aRMSE in Table \ref{tab:tab6}, and identifies spike-timing accuracy as a priority for further model development.

\begin{figure}[htbp]
	\centering
	\includegraphics[width=\textwidth]{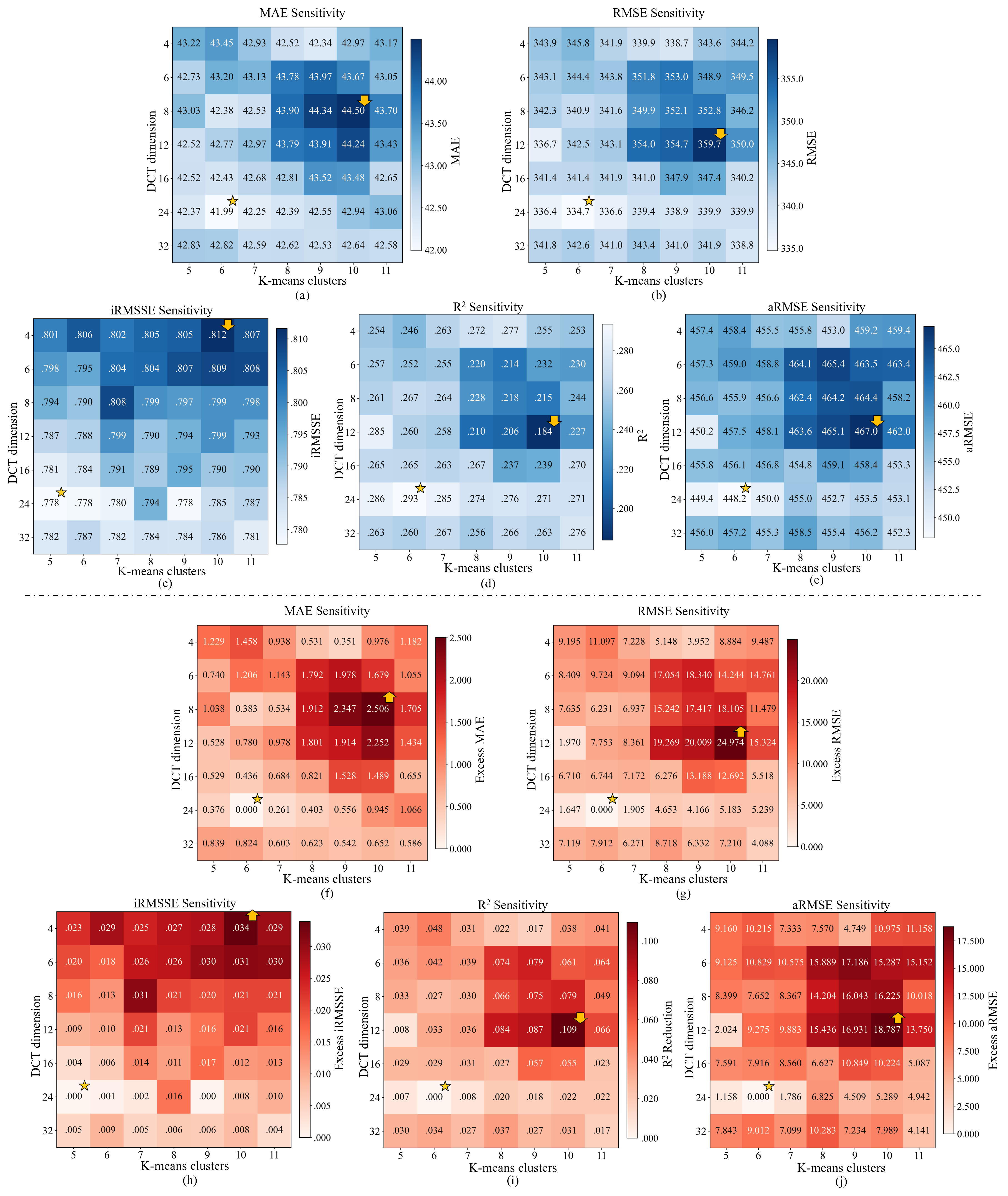}
	\caption{Sensitivity of CP-RAF at a 15-day horizon to the number of K-means clusters (x-axis) and the DCT coefficient dimension (y-axis). Panels (a)--(e) report MAE, RMSE, iRMSSE, \(\mathbf{R}^{\mathbf{2}}\), and aRMSE; panels (f)--(j) show the corresponding degradation relative to the best configuration. Stars and arrows mark the best and worst configurations, respectively.}
	\label{fig:fig6}
\end{figure}

Figure \ref{fig:fig6} shows whether the performance strongly depends on the number of K-means clusters and the DCT coefficient dimension for a 15-day horizon. The upper heatmaps show the MAE, RMSE, iRMSSE, \(R^{2}\), and aRMSE, and the lower heatmaps show the degradation relative to the best configuration for each metric.

Configurations with fewer than 16 DCT coefficients tend to perform worse, suggesting that a very low-dimensional representation removes the high-frequency features required to distinguish between abrupt increases and decreases. Nevertheless, the performance was relatively stable over the tested grid: the MAE ranges from 41.99 for the best configuration to 44.50 for the worst configuration, a difference of 2.51. The model is therefore not insensitive to the representation dimension, and its performance does not hinge on a single cluster--dimension pair. This moderate variation supports the robustness of CP-RAF within the evaluated range, while cautioning against overly aggressive DCT truncation.

\begin{figure}[htbp]
	\centering
	\includegraphics[width=\textwidth]{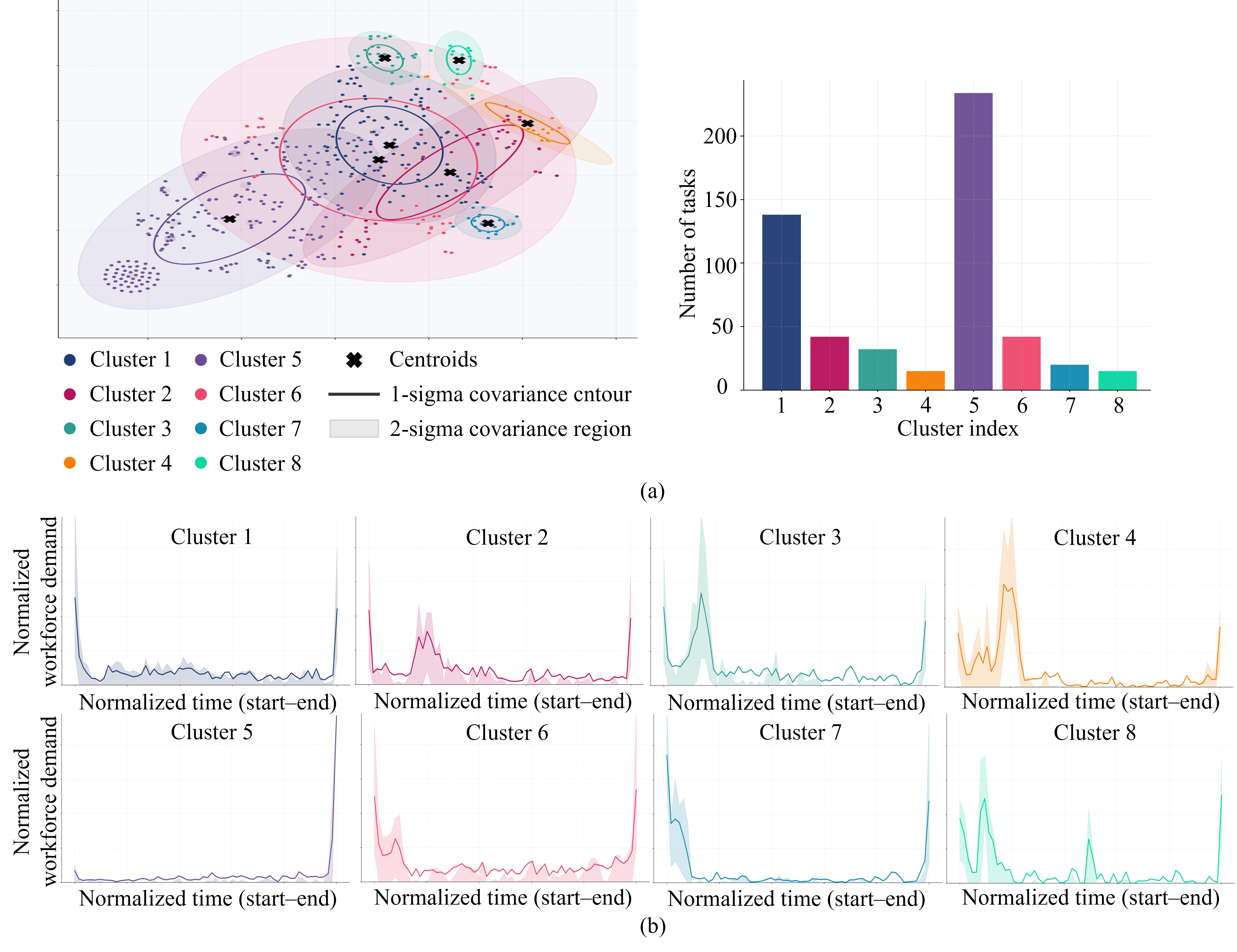}
	\caption{CP-RAF clustering results and normalized workforce-allocation profiles within each cluster.}
	\label{fig:fig7}
\end{figure}

Figure \ref{fig:fig7}(a) shows the number of tasks assigned to each cluster. An eight-cluster solution was selected for the validation set. Figure \ref{fig:fig7}(b) plots the normalized workforce-allocation profiles within each cluster. Cluster 1 is comparatively stable through the middle of a task but rises near the beginning and end. Clusters 2--4 shared an early-to-middle concentration pattern, with the peak becoming sharper from Cluster 2 to Cluster 4. Cluster 5 was predominantly late-concentrated. Clusters 6 and 7 resemble Cluster 1, but have sharper boundary peaks and lower mid-task allocations. Cluster 8 was distinguished by a modest increase during the mid-task. These clusters provide an empirical vocabulary for the recurring allocation shapes. Instead of imposing only coarse early-, middle-, and late-concentration categories, clustering distinguishes the degrees and combinations of concentrations present in the data. It also makes an individual forecast easier to trace. The target's observed prefix determines its cluster, and the prediction is assembled from the completed tasks within that cluster. This interpretation can help practitioners inspect the reference set underlying a forecast, and assess whether the retrieved patterns are operationally plausible.

Table \ref{tab:tab7} summarizes the online inference latencies of the methods compared. Measurements were repeated 100 times for each model, and the 50\textsuperscript{th}, 95\textsuperscript{th}, and 99\textsuperscript{th} percentiles were recorded. All experiments were conducted in Python 3.8 on a 64-bit Windows 11 workstation equipped with an Intel Core i9-14900KF CPU operating at 3.20 GHz. Classical statistical models, linear and tree-based models, and the proposed methods were executed on a CPU. The neural baselines were implemented in PyTorch and timed on a CPU. Unless otherwise stated, each measurement represents the elapsed time required to generate the seven-step forecasts for the 32 tasks.

\begin{longtable}[]{@{}
		>{\raggedright\arraybackslash}p{(\linewidth - 8\tabcolsep) * \real{0.2003}}
		>{\raggedright\arraybackslash}p{(\linewidth - 8\tabcolsep) * \real{0.2000}}
		>{\raggedright\arraybackslash}p{(\linewidth - 8\tabcolsep) * \real{0.2000}}
		>{\raggedright\arraybackslash}p{(\linewidth - 8\tabcolsep) * \real{0.2000}}
		>{\raggedright\arraybackslash}p{(\linewidth - 8\tabcolsep) * \real{0.1998}}@{}}
	\caption{Online batch-inference latency for a batch size of 32 and a forecast horizon of seven-time steps. The method-specific offline setup time is additionally reported for the proposed method.}\label{tab:tab7}\\
	\toprule\noalign{}
	\begin{minipage}[b]{\linewidth}\raggedright
		Model
	\end{minipage} & \begin{minipage}[b]{\linewidth}\raggedright
		Offline settings (s)
	\end{minipage} & \multicolumn{3}{l@{}}{Online Batch inference latency (s)} \\
	& & P50 & P95 & P99 \\
	\midrule\noalign{}
	\endfirsthead
	\toprule\noalign{}
	\begin{minipage}[b]{\linewidth}\raggedright
		Model
	\end{minipage} & \begin{minipage}[b]{\linewidth}\raggedright
		Offline settings (s)
	\end{minipage} & \multicolumn{3}{l@{}}{Online Batch inference latency (s)} \\
	& & P50 & P95 & P99 \\
	\midrule\noalign{}
	\endhead
	\bottomrule\noalign{}
	\endlastfoot
	Box--Jenkins & --- & 0.0197 & 0.0331 & 0.0357 \\
	LR & --- & 0.0005 & 0.0007 & 0.0009 \\
	SARIMAX & --- & 0.0838 & 0.0934 & 0.0971 \\
	LSTM & --- & 0.0015 & 0.0018 & 0.0020 \\
	XGBoost & --- & 0.0039 & 0.0048 & 0.0050 \\
	Hybrid LGBM & --- & 0.0049 & 0.0057 & 0.0061 \\
	NLinear & --- & 0.0002 & 0.0003 & 0.0004 \\
	RF & --- & 0.0159 & 0.0167 & 0.0170 \\
	Proposed & 37.8675 & 0.0339 & 0.0432 & 0.0446 \\
\end{longtable}

At the median, NLinear and LR exhibited the shortest latencies of 0.0002 and 0.0005 s, respectively. LSTM, XGBoost, and Hybrid LGBM also showed low median latencies of 0.0015, 0.0039, and 0.0049 s, respectively. Among the baselines, SARIMAX had the highest median latency at 0.0838 s, followed by Box--Jenkins at 0.0197 s and RF at 0.0159 s. The proposed method required 0.0339 s, which was slower than the other baselines, except for SARIMAX. The additional online cost arises from extracting the DCT coefficients of the target sequence, applying standardization, assigning the target to a fitted cluster, computing similarity weights, and aggregating the retrieved future trajectories. Nevertheless, the P99 latency of the proposed method was 0.0446 s, and all evaluated methods had P99 latencies below 0.1 s. Under the tested hardware and implementation, the computational time of the proposed method is therefore unlikely to constrain the daily workforce-planning cycle considered in this study.

In addition, the proposed method required 37.8675 s for the offline setup. This stage included extracting the DCT coefficients from the training and validation tasks, standardizing the coefficient vectors, fitting K-means using the number of clusters selected in the validation set, and constructing a reference future-trajectory database. The offline setup is not repeated for every forecast, but must be repeated whenever the reference database or the preprocessing and clustering configuration is updated. Therefore, the cost can be amortized over the online forecasts generated between database updates.

\begin{longtable}[]{@{}
		>{\raggedright\arraybackslash}p{(\linewidth - 8\tabcolsep) * \real{0.3221}}
		>{\raggedright\arraybackslash}p{(\linewidth - 8\tabcolsep) * \real{0.2484}}
		>{\raggedright\arraybackslash}p{(\linewidth - 8\tabcolsep) * \real{0.1433}}
		>{\raggedright\arraybackslash}p{(\linewidth - 8\tabcolsep) * \real{0.1433}}
		>{\raggedright\arraybackslash}p{(\linewidth - 8\tabcolsep) * \real{0.1430}}@{}}
	\caption{Offline setup time and online batch-inference latency of the proposed method as a function of reference-database size (batch size: 32; forecast horizon: seven-time steps).}\label{tab:tab8}\\
	\toprule\noalign{}
	\begin{minipage}[b]{\linewidth}\raggedright
		No. of retrieval tasks (K)
	\end{minipage} & \begin{minipage}[b]{\linewidth}\raggedright
		Offline settings (s)
	\end{minipage} & \multicolumn{3}{l@{}}{Online Batch inference latency (s)} \\
	& & P50 & P95 & P99 \\
	\midrule\noalign{}
	\endfirsthead
	\toprule\noalign{}
	\begin{minipage}[b]{\linewidth}\raggedright
		No. of retrieval tasks (K)
	\end{minipage} & \begin{minipage}[b]{\linewidth}\raggedright
		Offline settings (s)
	\end{minipage} & \multicolumn{3}{l@{}}{Online Batch inference latency (s)} \\
	& & P50 & P95 & P99 \\
	\midrule\noalign{}
	\endhead
	\bottomrule\noalign{}
	\endlastfoot
	653 (original experiment) & 37.8675 & 0.0339 & 0.0432 & 0.0446 \\
	1,000 & 37.8658 & 0.0354 & 0.0443 & 0.0462 \\
	10,000 & 37.8707 & 0.0351 & 0.0421 & 0.0572 \\
	100,000 & 37.8890 & 0.0376 & 0.0502 & 0.0547 \\
	1,000,000 & 38.1967 & 0.0820 & 0.1130 & 0.1265 \\
\end{longtable}

Table \ref{tab:tab8} lists the computational scaling of the proposed method as the number of tasks in the reference database increases. For this computational stress test, the original pool of training and validation tasks was expanded by generating Gaussian noise-perturbed replicas, resulting in reference databases containing 653, 1,000, 10,000, 100,000, and 1,000,000 tasks. Because the enlarged databases consisted of synthetic replicas rather than independently observed tasks, this experiment evaluated the computational latency only.

The median online latency remained similar between the 653 and 100,000 reference tasks, increasing from 0.0339 s to 0.0376 s. For the 1,000,000 tasks, the median latency increased to 0.0820 s, whereas the corresponding P50 and P99 latencies were 0.0820 and 0.1265 s, respectively. The offline setup time varied only marginally, from 37.8675 s with 653 reference tasks to 38.1967 s with 1,000,000 reference tasks. Under the evaluated hardware and timing protocol, the online retrieval and aggregation procedure therefore retained a P99 latency below 0.13 s with as many as one million synthetic reference tasks. This result supported the computational feasibility of the online forecasting stage under specified benchmark conditions.

\subsection{Sensitivity analysis}
\label{sec:sec4-5}

The proposed framework comprises two main stages: (i) representing the observed workforce-demand sequence in a feature space, and (ii) forecasting the remaining sequence by retrieving completed tasks with similar observed profiles and aggregating their future workforce-demand trajectories. The default configuration represents each observed sequence using DCT coefficients, and compares tasks using a cosine-based measure. Alternative representations include the raw workforce-demand sequence, discrete wavelet transform (DWT) coefficients, and Legendre-polynomial coefficients.

Two controlled comparisons were conducted to examine the sensitivity of the framework to design choices. First, the representation was fixed to the DCT coefficient vector while the retrieval measure was varied. Second, the cosine-based measure was fixed, whereas the sequence representation varied among the raw sequence, DWT coefficients, Legendre-polynomial coefficients, and DCT coefficients.

\begin{longtable}[]{@{}
		>{\raggedright\arraybackslash}p{(\linewidth - 10\tabcolsep) * \real{0.2895}}
		>{\raggedright\arraybackslash}p{(\linewidth - 10\tabcolsep) * \real{0.1512}}
		>{\raggedright\arraybackslash}p{(\linewidth - 10\tabcolsep) * \real{0.1538}}
		>{\raggedright\arraybackslash}p{(\linewidth - 10\tabcolsep) * \real{0.1406}}
		>{\raggedright\arraybackslash}p{(\linewidth - 10\tabcolsep) * \real{0.1474}}
		>{\raggedright\arraybackslash}p{(\linewidth - 10\tabcolsep) * \real{0.1175}}@{}}
	\caption{Forecasting performance obtained using four retrieval measures with the DCT coefficient representation. Each value is the mean of the corresponding metric across the nine progress ratios. The retrieval measure is used to compare tasks in Equations (\eqref{eq:5}) and (\eqref{eq:6}).}\label{tab:tab9}\\
	\toprule\noalign{}
	\begin{minipage}[b]{\linewidth}\raggedright
		Distance measure
	\end{minipage} & \multicolumn{5}{l@{}}{Avg. Performance for nine progress ratios} \\
	& MAE & RMSE & iRMSSE & \(R^{2}\) & aRMSE \\
	\midrule\noalign{}
	\endfirsthead
	\toprule\noalign{}
	\begin{minipage}[b]{\linewidth}\raggedright
		Distance measure
	\end{minipage} & \multicolumn{5}{l@{}}{Avg. Performance for nine progress ratios} \\
	& MAE & RMSE & iRMSSE & \(R^{2}\) & aRMSE \\
	\midrule\noalign{}
	\endhead
	\bottomrule\noalign{}
	\endlastfoot
	\textbf{Cosine} & \textbf{33.572} & \textbf{369.820} & \textbf{0.763} & \textbf{0.230} & \textbf{501.403} \\
	Manhattan & 34.599 & 381.180 & \underline{0.772} & \underline{0.182} & 508.322 \\
	Euclidean & 34.618 & 382.076 & 0.774 & 0.178 & 509.134 \\
	Mahalanobis & \underline{34.545} & \underline{380.044} & 0.779 & 0.178 & \underline{504.234} \\
\end{longtable}

Table \ref{tab:tab9} summarizes the mean values of the five evaluation metrics across nine progress ratios. The results for the individual progress ratios are presented in Table \ref{tab:appD1} in Appendix \ref{sec:appD}. The cosine-based measure produced the best numerical mean for all five metrics, yielding the lowest MAE, RMSE, iRMSSE, and aRMSE, and the highest \(R^{2}\).

The cosine-based measure evaluates the directional alignment between two non-zero vectors and is invariant to the positive rescaling of either vector. By contrast, the Manhattan and Euclidean distances are sensitive to absolute differences in the coefficient magnitude, whereas the Mahalanobis distance quantifies the coefficient differences relative to the estimated covariance structure. Therefore, the results suggest that, under the preprocessing and weighting scheme adopted in this study, the directional pattern of the DCT coefficients provides a more informative basis for retrieving reference tasks than their absolute or covariance-adjusted differences.

\begin{longtable}[]{@{}
		>{\raggedright\arraybackslash}p{(\linewidth - 10\tabcolsep) * \real{0.2895}}
		>{\raggedright\arraybackslash}p{(\linewidth - 10\tabcolsep) * \real{0.1512}}
		>{\raggedright\arraybackslash}p{(\linewidth - 10\tabcolsep) * \real{0.1538}}
		>{\raggedright\arraybackslash}p{(\linewidth - 10\tabcolsep) * \real{0.1406}}
		>{\raggedright\arraybackslash}p{(\linewidth - 10\tabcolsep) * \real{0.1474}}
		>{\raggedright\arraybackslash}p{(\linewidth - 10\tabcolsep) * \real{0.1175}}@{}}
	\caption{Forecasting performance obtained using four temporal-shape representations with the cosine-based retrieval measure in Equation (\eqref{eq:5}) held fixed. Each value is the mean of the corresponding metric across the nine progress ratios.}\label{tab:tab10}\\
	\toprule\noalign{}
	\begin{minipage}[b]{\linewidth}\raggedright
		Basis
	\end{minipage} & \multicolumn{5}{l@{}}{Avg. Performance for nine progress ratios} \\
	& MAE & RMSE & iRMSSE & \(R^{2}\) & aRMSE \\
	\midrule\noalign{}
	\endfirsthead
	\toprule\noalign{}
	\begin{minipage}[b]{\linewidth}\raggedright
		Basis
	\end{minipage} & \multicolumn{5}{l@{}}{Avg. Performance for nine progress ratios} \\
	& MAE & RMSE & iRMSSE & \(R^{2}\) & aRMSE \\
	\midrule\noalign{}
	\endhead
	\bottomrule\noalign{}
	\endlastfoot
	None (Raw signal) & 44.623 & 529.782 & 0.832 & -0.302 & 661.861 \\
	DWT & \underline{33.977} & \underline{374.933} & \textbf{0.760} & 0.198 & 505.711 \\
	Legendre & 34.095 & 377.027 & 0.772 & \underline{0.200} & \underline{504.853} \\
	\textbf{DCT} & \textbf{33.572} & \textbf{369.820} & \underline{0.763} & \textbf{0.230} & \textbf{501.403} \\
\end{longtable}

Table \ref{tab:tab10} compares the raw workforce-demand sequence with the DWT, Legendre-polynomial, and DCT coefficient representations. The reported values were averaged across nine progress ratios from 0.1 to 0.9, and the complete ratio-specific results are provided in Table \ref{tab:appD2} in Appendix \ref{sec:appD}.

Under the evaluated configuration, all three transformed representations substantially outperformed the raw-sequence representation across five metrics. For example, the raw representation produced a mean MAE of 44.623 and a negative mean \(R^{2}\) of -0.302, whereas the transformed representations produced mean MAEs between 33.572 and 34.095, with positive mean \(R^{2}\) values between 0.198 and 0.230. These findings indicate that the transformed coefficient representations used in this study support more effective reference-task retrieval than the raw-sequence representation implemented in the framework. The performance differences between DWT, Legendre, and DCT were considerably smaller than their respective differences from the raw representation. The DCT achieved the best mean MAE, RMSE, \(R^{2}\), and aRMSE, whereas the DWT achieved the lowest iRMSSE. Therefore, DCT did not dominate every evaluation metric, although it provided the best numerical results for four of the five metrics.

\section{Conclusion}
\label{sec:sec5}

Construction workforce planning requires forecasts that extend to the planned completion of each task, while remaining consistent with the workforce allocated in advance. This study formulated these requirements as a residual-allocation problem and developed a CP-RAF. The CP-RAF represents the observed segment of each task using DCT coefficients, retrieves completed tasks with similar demand shapes, and combines residual allocation profiles using similarity-based weights. Furthermore, the CRAA maps each reference profile to the remaining horizon of the target task while preserving the non-negativity and sum-to-one properties. Scaling the resulting profile using a known residual workforce produces a task-specific forecast, whose sum equals the planned residual demand. Therefore, these aggregate demand constraints are satisfied by construction rather than being imposed through post-hoc correction.

The evaluation used 93,915 daily observations of 770 tasks recorded over 447 days at a shipyard construction site. In the fixed-horizon experiments, CP-RAF achieved the best values for all five evaluation metrics on days 5, 7, 10, and 15. At the 3-day horizon, RF yielded a lower MAE and RMSE and a higher \(R^{2}\), whereas CP-RAF achieved the lowest iRMSSE and aRMSE. DM tests identified significant loss differences in 30 of the 40 model--horizon comparisons; all eight comparisons favored CP-RAF at the 7- and 10-day horizons at the 5\% level and at the 15-day horizon at the 0.5\% level. In the variable-horizon experiment, the CP-RAF generated forecasts for each task's planned completion for residual horizons of up to 291 days. The MAE remained between 32.741 and 35.156 for progress ratios from 0.1 to 0.9. Additional analyses showed that basis-derived representations performed substantially better than raw sequences, although the differences among the DCT, DWT, and Legendre representations were modest. Among the four distance measures examined, cosine similarity produced the best average performance. The median online inference time was 0.0339 s for 653 reference tasks, which was shorter than the daily planning cycle considered in this study.

This study was limited to a univariate DCT-based representation and cluster-restricted reference retrieval. Future work can extend the CP-RAF to multivariate inputs using dimensionality reduction methods or neural encoders, thereby providing richer latent representations for identifying relevant residual allocation patterns. Explicit anomaly detection prior to clustering may also reduce the influence of atypical tasks characterized by unusual scales, temporal dynamics, or cross-variable dependencies. Another approach involves an end-to-end forecasting model that incorporates aggregate demand requirements as a constraint-aware inductive bias. These research directions can broaden the practical applicability of the proposed framework while accommodating the operational constraints encountered at construction sites.

\section*{Acknowledgements}

This work was supported by the National Research Foundation of Korea(NRF) grant funded by the Korea government(MSIT) (No.RS-2023-00218913).

\bibliographystyle{plainnat}
\bibliography{references}

\appendix
\renewcommand{\thetable}{\thesection\Roman{table}}
\renewcommand{\theequation}{\thesection.\arabic{equation}}
\renewcommand{\theHtable}{\thesection.\arabic{table}}
\renewcommand{\theHequation}{\thesection.\arabic{equation}}
\section{Computational detail for CRAA}
\label{sec:appA}
\setcounter{table}{0}
\setcounter{equation}{0}

Let Piecewise-linear cumulative evaluation,

\begin{equation}
	\label{eq:A-1}
	S_{q,0} = 0,\ \ S_{q,h} = \sum_{b = 1}^{h}S_{q,h},\ \ h = 1,\ldots,L_{q}
\end{equation}

For \(0 \leq v < 1\), set \(x = vL_{q}\), \(b = \left\lfloor x \right\rfloor\), and \(\gamma = x - b\). The cumulative allocation curves in Equation (\refeq{eq:9}) can be evaluated as follows:

\begin{equation}
	\label{eq:A-2}
	A_{q}(v) = (1 - \gamma)S_{q,b} + \gamma S_{q,b + 1},\ \ A_{q}(1) = 1
\end{equation}

\begin{longtable}[]{@{}
		>{\raggedright\arraybackslash}p{(\linewidth - 4\tabcolsep) * \real{0.0524}}
		>{\raggedright\arraybackslash}p{(\linewidth - 4\tabcolsep) * \real{0.0470}}
		>{\raggedright\arraybackslash}p{(\linewidth - 4\tabcolsep) * \real{0.9006}}@{}}
	\toprule\noalign{}
	\multicolumn{3}{@{}l@{}}{\textbf{Algorithm 1} Cumulative Residual Allocation Alignment (CRAA)} \\
	\midrule\noalign{}
	\endhead
	\bottomrule\noalign{}
	\endlastfoot
	\multicolumn{3}{@{}l@{}}{%
		\textbf{Input:} normalized reference allocation \(\mathbf{a}_{q} = \left( a_{q,1},\ldots,a_{q,L_{q}} \right)^{T}\); target residual horizon \(L_{i}\).} \\
	\multicolumn{3}{@{}l@{}}{%
		\textbf{Output:} target-aligned residual allocation \({\overline{\mathbf{a}}}_{q \rightarrow i} = \left\{ {\overline{a}}_{q \rightarrow i,1},\ldots,{\overline{a}}_{q \rightarrow i,L_{i}} \right\} \in \mathbb{R}_{+}^{L_{i}}\).} \\
	1: & \multicolumn{2}{l@{}}{%
		\(S_{0} \leftarrow 0\)} \\
	2: & \multicolumn{2}{l@{}}{%
		\textbf{for} \(h = 1,\ldots,L_{q}\) \textbf{do}} \\
	3: & & \(S_{q,h} \leftarrow S_{q,h - 1} + a_{q,h}\) \\
	4: & \multicolumn{2}{l@{}}{%
		\textbf{end for}} \\
	5: & \multicolumn{2}{l@{}}{%
		\textbf{for} \(j = 1,\ldots,L_{i}\) \textbf{do}} \\
	6: & & \(v_{j}^{-} \leftarrow \frac{(j - 1)}{L_{i}}\) \\
	7: & & \(v_{j}^{+} \leftarrow \frac{j}{L_{i}}\) \\
	8: & & \emph{Evaluate} \(A_{q}\left( v_{j}^{-} \right)\) \emph{and} \(A_{q}\left( v_{j}^{+} \right)\) \emph{using Eq. (A.2)} \\
	9: & & \({\overline{\alpha}}_{q \rightarrow i,j} \leftarrow A_{q}\left( v_{j}^{+} \right) - A_{q}\left( v_{j}^{-} \right)\) \\
	10: & \multicolumn{2}{l@{}}{%
		\textbf{end for}} \\
	11: & \multicolumn{2}{l@{}}{%
		\textbf{return} \({\overline{\mathbf{a}}}_{q \rightarrow i} = \left\{ {\overline{a}}_{q \rightarrow i,1},\ldots,{\overline{a}}_{q \rightarrow i,L_{i}} \right\}\)} \\
\end{longtable}

\section{Supplementary dataset information}
\label{sec:appB}
\setcounter{table}{0}
\setcounter{equation}{0}

\begin{longtable}[]{@{}
		>{\raggedright\arraybackslash}p{(\linewidth - 20\tabcolsep) * \real{0.1151}}
		>{\raggedright\arraybackslash}p{(\linewidth - 20\tabcolsep) * \real{0.1135}}
		>{\raggedright\arraybackslash}p{(\linewidth - 20\tabcolsep) * \real{0.1390}}
		>{\raggedright\arraybackslash}p{(\linewidth - 20\tabcolsep) * \real{0.0849}}
		>{\raggedright\arraybackslash}p{(\linewidth - 20\tabcolsep) * \real{0.0738}}
		>{\raggedright\arraybackslash}p{(\linewidth - 20\tabcolsep) * \real{0.0609}}
		>{\raggedright\arraybackslash}p{(\linewidth - 20\tabcolsep) * \real{0.0646}}
		>{\raggedright\arraybackslash}p{(\linewidth - 20\tabcolsep) * \real{0.0622}}
		>{\raggedright\arraybackslash}p{(\linewidth - 20\tabcolsep) * \real{0.0622}}
		>{\raggedright\arraybackslash}p{(\linewidth - 20\tabcolsep) * \real{0.0622}}
		>{\raggedright\arraybackslash}p{(\linewidth - 20\tabcolsep) * \real{0.1618}}@{}}
	\caption{Composition and time-series length statistics by data split.}\label{tab:appB1}\\
	\toprule\noalign{}
	\begin{minipage}[b]{\linewidth}\raggedright
		Split
	\end{minipage} & \begin{minipage}[b]{\linewidth}\raggedright
		No. of tasks
	\end{minipage} & \begin{minipage}[b]{\linewidth}\raggedright
		No. of samples
	\end{minipage} & \begin{minipage}[b]{\linewidth}\raggedright
		Mean
	\end{minipage} & \begin{minipage}[b]{\linewidth}\raggedright
		SD
	\end{minipage} & \begin{minipage}[b]{\linewidth}\raggedright
		Min
	\end{minipage} & \begin{minipage}[b]{\linewidth}\raggedright
		Max
	\end{minipage} & \begin{minipage}[b]{\linewidth}\raggedright
		Q25
	\end{minipage} & \begin{minipage}[b]{\linewidth}\raggedright
		Q50
	\end{minipage} & \begin{minipage}[b]{\linewidth}\raggedright
		Q75
	\end{minipage} & \begin{minipage}[b]{\linewidth}\raggedright
		Zero samples
	\end{minipage} \\
	\midrule\noalign{}
	\endfirsthead
	\toprule\noalign{}
	\begin{minipage}[b]{\linewidth}\raggedright
		Split
	\end{minipage} & \begin{minipage}[b]{\linewidth}\raggedright
		No. of tasks
	\end{minipage} & \begin{minipage}[b]{\linewidth}\raggedright
		No. of samples
	\end{minipage} & \begin{minipage}[b]{\linewidth}\raggedright
		Mean
	\end{minipage} & \begin{minipage}[b]{\linewidth}\raggedright
		SD
	\end{minipage} & \begin{minipage}[b]{\linewidth}\raggedright
		Min
	\end{minipage} & \begin{minipage}[b]{\linewidth}\raggedright
		Max
	\end{minipage} & \begin{minipage}[b]{\linewidth}\raggedright
		Q25
	\end{minipage} & \begin{minipage}[b]{\linewidth}\raggedright
		Q50
	\end{minipage} & \begin{minipage}[b]{\linewidth}\raggedright
		Q75
	\end{minipage} & \begin{minipage}[b]{\linewidth}\raggedright
		Zero samples
	\end{minipage} \\
	\midrule\noalign{}
	\endhead
	\bottomrule\noalign{}
	\endlastfoot
	Train & 538 & 65,564 & 121.87 & 65.52 & 11 & 378 & 70 & 112 & 164 & 47,483 \\
	Validation & 115 & 14,000 & 121.74 & 65.77 & 16 & 295 & 70 & 117 & 154 & 9,985 \\
	Test & 117 & 14,351 & 122.66 & 65.09 & 11 & 323 & 80 & 110 & 163 & 9,963 \\
\end{longtable}

\begin{longtable}[]{@{}
		>{\raggedright\arraybackslash}p{(\linewidth - 20\tabcolsep) * \real{0.1151}}
		>{\raggedright\arraybackslash}p{(\linewidth - 20\tabcolsep) * \real{0.1149}}
		>{\raggedright\arraybackslash}p{(\linewidth - 20\tabcolsep) * \real{0.1408}}
		>{\raggedright\arraybackslash}p{(\linewidth - 20\tabcolsep) * \real{0.0745}}
		>{\raggedright\arraybackslash}p{(\linewidth - 20\tabcolsep) * \real{0.0849}}
		>{\raggedright\arraybackslash}p{(\linewidth - 20\tabcolsep) * \real{0.0627}}
		>{\raggedright\arraybackslash}p{(\linewidth - 20\tabcolsep) * \real{0.0849}}
		>{\raggedright\arraybackslash}p{(\linewidth - 20\tabcolsep) * \real{0.0627}}
		>{\raggedright\arraybackslash}p{(\linewidth - 20\tabcolsep) * \real{0.0627}}
		>{\raggedright\arraybackslash}p{(\linewidth - 20\tabcolsep) * \real{0.0627}}
		>{\raggedright\arraybackslash}p{(\linewidth - 20\tabcolsep) * \real{0.1342}}@{}}
	\caption{Distributional statistics of the daily workforce demand by data split.}\label{tab:appB2}\\
	\toprule\noalign{}
	\begin{minipage}[b]{\linewidth}\raggedright
		Split
	\end{minipage} & \begin{minipage}[b]{\linewidth}\raggedright
		No. of tasks
	\end{minipage} & \begin{minipage}[b]{\linewidth}\raggedright
		No. of samples
	\end{minipage} & \begin{minipage}[b]{\linewidth}\raggedright
		Mean
	\end{minipage} & \begin{minipage}[b]{\linewidth}\raggedright
		SD
	\end{minipage} & \begin{minipage}[b]{\linewidth}\raggedright
		Min
	\end{minipage} & \begin{minipage}[b]{\linewidth}\raggedright
		Max
	\end{minipage} & \begin{minipage}[b]{\linewidth}\raggedright
		Q25
	\end{minipage} & \begin{minipage}[b]{\linewidth}\raggedright
		Q50
	\end{minipage} & \begin{minipage}[b]{\linewidth}\raggedright
		Q75
	\end{minipage} & \begin{minipage}[b]{\linewidth}\raggedright
		Zero ratio (\%)
	\end{minipage} \\
	\midrule\noalign{}
	\endfirsthead
	\toprule\noalign{}
	\begin{minipage}[b]{\linewidth}\raggedright
		Split
	\end{minipage} & \begin{minipage}[b]{\linewidth}\raggedright
		No. of tasks
	\end{minipage} & \begin{minipage}[b]{\linewidth}\raggedright
		No. of samples
	\end{minipage} & \begin{minipage}[b]{\linewidth}\raggedright
		Mean
	\end{minipage} & \begin{minipage}[b]{\linewidth}\raggedright
		SD
	\end{minipage} & \begin{minipage}[b]{\linewidth}\raggedright
		Min
	\end{minipage} & \begin{minipage}[b]{\linewidth}\raggedright
		Max
	\end{minipage} & \begin{minipage}[b]{\linewidth}\raggedright
		Q25
	\end{minipage} & \begin{minipage}[b]{\linewidth}\raggedright
		Q50
	\end{minipage} & \begin{minipage}[b]{\linewidth}\raggedright
		Q75
	\end{minipage} & \begin{minipage}[b]{\linewidth}\raggedright
		Zero ratio (\%)
	\end{minipage} \\
	\midrule\noalign{}
	\endhead
	\bottomrule\noalign{}
	\endlastfoot
	Train & 538 & 65,564 & 26.01 & 175.75 & 0.00 & 16,380 & 0.00 & 0.00 & 1.17 & 72.42 \\
	Validation & 115 & 14,000 & 39.40 & 266.95 & 0.00 & 12,511 & 0.00 & 0.00 & 1.17 & 71.32 \\
	Test & 117 & 14,351 & 42.09 & 377.79 & 0.00 & & 0.00 & 0.00 & 2.34 & 69.42 \\
\end{longtable}

\begin{longtable}[]{@{}
		>{\raggedright\arraybackslash}p{(\linewidth - 14\tabcolsep) * \real{0.1992}}
		>{\raggedright\arraybackslash}p{(\linewidth - 14\tabcolsep) * \real{0.2752}}
		>{\raggedright\arraybackslash}p{(\linewidth - 14\tabcolsep) * \real{0.1012}}
		>{\raggedright\arraybackslash}p{(\linewidth - 14\tabcolsep) * \real{0.0828}}
		>{\raggedright\arraybackslash}p{(\linewidth - 14\tabcolsep) * \real{0.0877}}
		>{\raggedright\arraybackslash}p{(\linewidth - 14\tabcolsep) * \real{0.0846}}
		>{\raggedright\arraybackslash}p{(\linewidth - 14\tabcolsep) * \real{0.0846}}
		>{\raggedright\arraybackslash}p{(\linewidth - 14\tabcolsep) * \real{0.0846}}@{}}
	\caption{Summary of eligible test tasks and remaining forecast horizons by progress ratio.}\label{tab:appB3}\\
	\toprule\noalign{}
	\begin{minipage}[b]{\linewidth}\raggedright
		Progress ratio
	\end{minipage} & \begin{minipage}[b]{\linewidth}\raggedright
		No. of eligible tasks
	\end{minipage} & \begin{minipage}[b]{\linewidth}\raggedright
		Mean
	\end{minipage} & \begin{minipage}[b]{\linewidth}\raggedright
		Min
	\end{minipage} & \begin{minipage}[b]{\linewidth}\raggedright
		Max
	\end{minipage} & \begin{minipage}[b]{\linewidth}\raggedright
		Q25
	\end{minipage} & \begin{minipage}[b]{\linewidth}\raggedright
		Q50
	\end{minipage} & \begin{minipage}[b]{\linewidth}\raggedright
		Q75
	\end{minipage} \\
	\midrule\noalign{}
	\endfirsthead
	\toprule\noalign{}
	\begin{minipage}[b]{\linewidth}\raggedright
		Progress ratio
	\end{minipage} & \begin{minipage}[b]{\linewidth}\raggedright
		No. of eligible tasks
	\end{minipage} & \begin{minipage}[b]{\linewidth}\raggedright
		Mean
	\end{minipage} & \begin{minipage}[b]{\linewidth}\raggedright
		Min
	\end{minipage} & \begin{minipage}[b]{\linewidth}\raggedright
		Max
	\end{minipage} & \begin{minipage}[b]{\linewidth}\raggedright
		Q25
	\end{minipage} & \begin{minipage}[b]{\linewidth}\raggedright
		Q50
	\end{minipage} & \begin{minipage}[b]{\linewidth}\raggedright
		Q75
	\end{minipage} \\
	\midrule\noalign{}
	\endhead
	\bottomrule\noalign{}
	\endlastfoot
	0.1 & 114 & 76 & 7 & 291 & 34 & 64 & 105 \\
	0.2 & 117 & 68 & 7 & 259 & 31 & 57 & 94 \\
	0.3 & 117 & 60 & 7 & 227 & 27 & 51 & 83 \\
	0.4 & 117 & 52 & 7 & 194 & 24 & 44 & 71 \\
	0.5 & 116 & 44 & 7 & 162 & 21 & 38 & 60 \\
	0.6 & 116 & 36 & 7 & 130 & 18 & 31 & 49 \\
	0.7 & 114 & 28 & 7 & 97 & 15 & 24 & 38 \\
	0.8 & 109 & 20 & 7 & 65 & 12 & 18 & 27 \\
	0.9 & 99 & 13 & 7 & 33 & 9 & 11 & 15 \\
\end{longtable}

\clearpage
\begin{landscape}
	\setlength{\manuscripttablewidth}{\linewidth}
	\setlength{\tabcolsep}{1pt}
	\renewcommand{\manuscripttablefont}{\fontsize{7.5}{9}\selectfont}
	\section{Selected hyperparameter configurations}
	\label{sec:appC}
	\setcounter{table}{0}
	\setcounter{equation}{0}
	
	\begin{longtable}[]{@{}
			>{\raggedright\arraybackslash}p{(\linewidth - 12\tabcolsep) * \real{0.1169}}
			>{\raggedright\arraybackslash}p{(\linewidth - 12\tabcolsep) * \real{0.2231}}
			>{\raggedright\arraybackslash}p{(\linewidth - 12\tabcolsep) * \real{0.1320}}
			>{\raggedright\arraybackslash}p{(\linewidth - 12\tabcolsep) * \real{0.1320}}
			>{\raggedright\arraybackslash}p{(\linewidth - 12\tabcolsep) * \real{0.1320}}
			>{\raggedright\arraybackslash}p{(\linewidth - 12\tabcolsep) * \real{0.1320}}
			>{\raggedright\arraybackslash}p{(\linewidth - 12\tabcolsep) * \real{0.1320}}@{}}
		\caption{Selected hyperparameter configurations for Exp-case 1.}\label{tab:tabC1}\\
		\toprule\noalign{}
		\begin{minipage}[b]{\linewidth}\raggedright
			Model
		\end{minipage} & \begin{minipage}[b]{\linewidth}\raggedright
			Hyperparameter combination
		\end{minipage} & \multicolumn{5}{l@{}}{Prediction horizon} \\
		& & 3 & 5 & 7 & 10 & 15 \\
		\midrule\noalign{}
		\endfirsthead
		\toprule\noalign{}
		\begin{minipage}[b]{\linewidth}\raggedright
			Model
		\end{minipage} & \begin{minipage}[b]{\linewidth}\raggedright
			Hyperparameter combination
		\end{minipage} & \multicolumn{5}{l@{}}{Prediction horizon} \\
		& & 3 & 5 & 7 & 10 & 15 \\
		\midrule\noalign{}
		\endhead
		\bottomrule\noalign{}
		\endlastfoot
		Box--Jenkins & \((p,q,d)\) & (5,1,0) & (5,1,0) & (3,1,0) & (0,0,3) & (0,0,3) \\
		SARIMAX & \((p,q,d,P,Q,D,s)\) & (2,0,1,0,1,1,7) & (1,0,2,1,1,2,7) & (0,0,3,2,1,1,7) & (0,0,0,2,1,1,7) & (0,0,0,2,1,1,7) \\
		LSTM & \(\left( h,L,r_{drop},\eta,\lambda_{reg},B \right)\) & (256,2,0.1,$10^{-4}$,$10^{-6}$,32) & (256,2,0.1,$10^{-3}$,$10^{-5}$,128) & (256,2,0.1,$10^{-4}$,$10^{-6}$,32) & (64,1,0.0,$10^{-3}$,$10^{-6}$,32) & (32,2,0.0,$10^{-3}$,$10^{-5}$,32) \\
		XGBoost & \(\left( N_{est},d_{\max},p_{col},w_{\min}^{child},\gamma,\alpha,\lambda \right)\) & (300,6,0.84,3,1,$10^{-2}$,2) & (300,8,0.83,10,0.1,$10^{-3}$,5) & (700,4,0.88,1,1,0.1,3) & (200,5,0.69,10,0.1,0.1,5) & (700,8,0.84,10,0.01,0,2) \\
		Hybrid LGBM & \(\left( N_{est},N_{leaf},d_{\max},n_{\min}^{child},p_{col},\alpha,\lambda \right)\) & (300,15,10,80,0.85,0,2) & (300,31,5,80,0.77,$10^{-4}$,5) & (300,63,5,80,0.77,0.1,2) & (200,31,10,10,0.78,0.1,2) & (300,15,10,40,0.96,0.1,0.5) \\
		NLinear & \(\left( \eta,\lambda_{reg},B,N_{est},d_{\max},n_{\min}^{split} \right)\) & ($10^{-3}$,$10^{-6}$,256) & ($10^{-4}$,$10^{-6}$,64) & ($10^{-3}$,$10^{-4}$,64) & ($10^{-3}$,$10^{-4}$,64) & ($10^{-4}$,$10^{-6}$,256) \\
		RF & \(\left( n_{\min}^{leaf},p_{col} \right)\) & (200,10,5,4,0.5) & (200,15,10,4,0.5) & (1000,15,5,4,0.8) & (200,10,2,4,0.8) & (1000,10,5,4,0.5) \\
		Proposed & \(\left( k,d_{dct} \right)\) & (11,12) & (10,12) & (11,12) & (8,24) & (6,24) \\
	\end{longtable}
	
	\begin{longtable}[]{@{}
			>{\raggedright\arraybackslash}p{(\linewidth - 18\tabcolsep) * \real{0.2691}}
			>{\raggedright\arraybackslash}p{(\linewidth - 18\tabcolsep) * \real{0.0838}}
			>{\raggedright\arraybackslash}p{(\linewidth - 18\tabcolsep) * \real{0.0838}}
			>{\raggedright\arraybackslash}p{(\linewidth - 18\tabcolsep) * \real{0.0830}}
			>{\raggedright\arraybackslash}p{(\linewidth - 18\tabcolsep) * \real{0.0838}}
			>{\raggedright\arraybackslash}p{(\linewidth - 18\tabcolsep) * \real{0.0734}}
			>{\raggedright\arraybackslash}p{(\linewidth - 18\tabcolsep) * \real{0.0830}}
			>{\raggedright\arraybackslash}p{(\linewidth - 18\tabcolsep) * \real{0.0734}}
			>{\raggedright\arraybackslash}p{(\linewidth - 18\tabcolsep) * \real{0.0830}}
			>{\raggedright\arraybackslash}p{(\linewidth - 18\tabcolsep) * \real{0.0838}}@{}}
		\caption{Selected hyperparameter configurations for Exp-case 2.}\label{tab:tabC2}\\
		\toprule\noalign{}
		\begin{minipage}[b]{\linewidth}\raggedright
			Hyperparameter combination
		\end{minipage} & \multicolumn{9}{l@{}}{Progress ratio} \\
		& 0.1 & 0.2 & 0.3 & 0.4 & 0.5 & 0.6 & 0.7 & 0.8 & 0.9 \\
		\midrule\noalign{}
		\endfirsthead
		\toprule\noalign{}
		\begin{minipage}[b]{\linewidth}\raggedright
			Hyperparameter combination
		\end{minipage} & \multicolumn{9}{l@{}}{Progress ratio} \\
		& 0.1 & 0.2 & 0.3 & 0.4 & 0.5 & 0.6 & 0.7 & 0.8 & 0.9 \\
		\midrule\noalign{}
		\endhead
		\bottomrule\noalign{}
		\endlastfoot
		\(\left( k,d_{dct} \right)\) & (10,24) & (10,24) & (11,32) & (10,24) & (8,24) & (11,24) & (8,24) & (11,32) & (10,24) \\
	\end{longtable}
	
\end{landscape}
\clearpage
\setlength{\manuscripttablewidth}{\textwidth}
\setlength{\tabcolsep}{4pt}
\renewcommand{\manuscripttablefont}{\footnotesize}
\section{Supplementary results for alternative distance measures and basis functions}
\label{sec:appD}
\setcounter{table}{0}
\setcounter{equation}{0}

\begin{longtable}[]{@{}
		>{\raggedright\arraybackslash}p{(\linewidth - 12\tabcolsep) * \real{0.2449}}
		>{\raggedright\arraybackslash}p{(\linewidth - 12\tabcolsep) * \real{0.1877}}
		>{\raggedright\arraybackslash}p{(\linewidth - 12\tabcolsep) * \real{0.1086}}
		>{\raggedright\arraybackslash}p{(\linewidth - 12\tabcolsep) * \real{0.1232}}
		>{\raggedright\arraybackslash}p{(\linewidth - 12\tabcolsep) * \real{0.1286}}
		>{\raggedright\arraybackslash}p{(\linewidth - 12\tabcolsep) * \real{0.1036}}
		>{\raggedright\arraybackslash}p{(\linewidth - 12\tabcolsep) * \real{0.1034}}@{}}
	\caption{Workforce demand forecasting performance across nine progress ratios using four distance measures.}\label{tab:appD1}\\
	\toprule\noalign{}
	\begin{minipage}[b]{\linewidth}\raggedright
		Distance measure
	\end{minipage} & \begin{minipage}[b]{\linewidth}\raggedright
		Progress ratio
	\end{minipage} & \begin{minipage}[b]{\linewidth}\raggedright
		MAE
	\end{minipage} & \begin{minipage}[b]{\linewidth}\raggedright
		RMSE
	\end{minipage} & \begin{minipage}[b]{\linewidth}\raggedright
		iRMSSE
	\end{minipage} & \begin{minipage}[b]{\linewidth}\raggedright
		R\textsuperscript{2}
	\end{minipage} & \begin{minipage}[b]{\linewidth}\raggedright
		aRMSE
	\end{minipage} \\
	\midrule\noalign{}
	\endfirsthead
	\toprule\noalign{}
	\begin{minipage}[b]{\linewidth}\raggedright
		Distance measure
	\end{minipage} & \begin{minipage}[b]{\linewidth}\raggedright
		Progress ratio
	\end{minipage} & \begin{minipage}[b]{\linewidth}\raggedright
		MAE
	\end{minipage} & \begin{minipage}[b]{\linewidth}\raggedright
		RMSE
	\end{minipage} & \begin{minipage}[b]{\linewidth}\raggedright
		iRMSSE
	\end{minipage} & \begin{minipage}[b]{\linewidth}\raggedright
		R\textsuperscript{2}
	\end{minipage} & \begin{minipage}[b]{\linewidth}\raggedright
		aRMSE
	\end{minipage} \\
	\midrule\noalign{}
	\endhead
	\bottomrule\noalign{}
	\endlastfoot
	Cosine & 0.1 & 32.877 & 313.104 & 0.795 & 0.243 & 420.794 \\
	& 0.2 & 32.823 & 321.144 & 0.792 & 0.248 & 432.058 \\
	& 0.3 & 33.173 & 326.694 & 0.788 & 0.260 & 439.919 \\
	& 0.4 & 33.404 & 341.165 & 0.779 & 0.245 & 458.309 \\
	& 0.5 & 34.383 & 356.587 & 0.783 & 0.226 & 478.401 \\
	& 0.6 & 33.597 & 358.875 & 0.768 & 0.272 & 486.128 \\
	& 0.7 & 34.001 & 404.248 & 0.748 & 0.164 & 542.202 \\
	& 0.8 & 35.156 & 437.848 & 0.728 & 0.210 & 591.559 \\
	& 0.9 & 32.741 & 468.723 & 0.686 & 0.202 & 663.260 \\
	\(\mu\) & & 33.572 & 369.820 & 0.763 & 0.230 & 501.403 \\ \midrule\noalign{}
	Manhattan & 0.1 & 34.335 & 324.285 & 0.815 & 0.189 & 430.445 \\
	& 0.2 & 34.427 & 332.744 & 0.804 & 0.193 & 442.146 \\
	& 0.3 & 34.383 & 341.321 & 0.799 & 0.193 & 453.807 \\
	& 0.4 & 34.537 & 352.462 & 0.793 & 0.195 & 468.863 \\
	& 0.5 & 34.813 & 365.162 & 0.786 & 0.188 & 486.197 \\
	& 0.6 & 34.896 & 376.389 & 0.775 & 0.200 & 503.736 \\
	& 0.7 & 35.007 & 402.994 & 0.763 & 0.170 & 539.264 \\
	& 0.8 & 35.914 & 449.353 & 0.730 & 0.168 & 602.025 \\
	& 0.9 & 33.087 & 485.910 & 0.689 & 0.143 & 648.416 \\
	\(\mu\) & & 34.599 & 381.180 & 0.772 & 0.182 & 508.322 \\ \midrule\noalign{}
	Euclidean & 0.1 & 34.302 & 325.351 & 0.815 & 0.183 & 431.454 \\
	& 0.2 & 34.353 & 333.034 & 0.805 & 0.191 & 442.480 \\
	& 0.3 & 34.467 & 343.545 & 0.803 & 0.183 & 455.870 \\
	& 0.4 & 34.425 & 351.272 & 0.788 & 0.201 & 467.806 \\
	& 0.5 & 34.845 & 365.739 & 0.789 & 0.186 & 486.545 \\
	& 0.6 & 34.864 & 376.409 & 0.777 & 0.200 & 503.616 \\
	& 0.7 & 35.165 & 405.197 & 0.767 & 0.161 & 541.317 \\
	& 0.8 & 35.956 & 451.070 & 0.733 & 0.162 & 603.578 \\
	& 0.9 & 33.187 & 487.068 & 0.690 & 0.139 & 649.547 \\
	\(\mu\) & & 34.618 & 382.076 & 0.774 & 0.178 & 509.134 \\ \midrule\noalign{}
	Mahalanobis & 0.1 & 34.652 & 329.248 & 0.850 & 0.164 & 434.394 \\
	& 0.2 & 34.649 & 339.377 & 0.823 & 0.160 & 448.498 \\
	& 0.3 & 35.408 & 353.959 & 0.813 & 0.132 & 466.030 \\
	& 0.4 & 34.341 & 369.197 & 0.781 & 0.117 & 483.634 \\
	& 0.5 & 35.635 & 402.603 & 0.799 & 0.013 & 519.901 \\
	& 0.6 & 34.214 & 362.734 & 0.789 & 0.257 & 486.081 \\
	& 0.7 & 34.171 & 370.364 & 0.747 & 0.299 & 489.907 \\
	& 0.8 & 34.767 & 433.724 & 0.718 & 0.225 & 586.243 \\
	& 0.9 & 33.073 & 459.190 & 0.691 & 0.235 & 623.419 \\
	\(\mu\) & & 34.545 & 380.044 & 0.779 & 0.178 & 504.234 \\
\end{longtable}

\begin{longtable}[]{@{}
		>{\raggedright\arraybackslash}p{(\linewidth - 12\tabcolsep) * \real{0.2450}}
		>{\raggedright\arraybackslash}p{(\linewidth - 12\tabcolsep) * \real{0.1876}}
		>{\raggedright\arraybackslash}p{(\linewidth - 12\tabcolsep) * \real{0.1086}}
		>{\raggedright\arraybackslash}p{(\linewidth - 12\tabcolsep) * \real{0.1230}}
		>{\raggedright\arraybackslash}p{(\linewidth - 12\tabcolsep) * \real{0.1286}}
		>{\raggedright\arraybackslash}p{(\linewidth - 12\tabcolsep) * \real{0.1036}}
		>{\raggedright\arraybackslash}p{(\linewidth - 12\tabcolsep) * \real{0.1036}}@{}}
	\caption{Workforce demand forecasting performance across nine progress ratios using four basis functions.}\label{tab:appD2}\\
	\toprule\noalign{}
	\begin{minipage}[b]{\linewidth}\raggedright
		Basis
	\end{minipage} & \begin{minipage}[b]{\linewidth}\raggedright
		Progress ratio
	\end{minipage} & \begin{minipage}[b]{\linewidth}\raggedright
		MAE
	\end{minipage} & \begin{minipage}[b]{\linewidth}\raggedright
		RMSE
	\end{minipage} & \begin{minipage}[b]{\linewidth}\raggedright
		iRMSSE
	\end{minipage} & \begin{minipage}[b]{\linewidth}\raggedright
		R\textsuperscript{2}
	\end{minipage} & \begin{minipage}[b]{\linewidth}\raggedright
		aRMSE
	\end{minipage} \\
	\midrule\noalign{}
	\endfirsthead
	\toprule\noalign{}
	\begin{minipage}[b]{\linewidth}\raggedright
		Basis
	\end{minipage} & \begin{minipage}[b]{\linewidth}\raggedright
		Progress ratio
	\end{minipage} & \begin{minipage}[b]{\linewidth}\raggedright
		MAE
	\end{minipage} & \begin{minipage}[b]{\linewidth}\raggedright
		RMSE
	\end{minipage} & \begin{minipage}[b]{\linewidth}\raggedright
		iRMSSE
	\end{minipage} & \begin{minipage}[b]{\linewidth}\raggedright
		R\textsuperscript{2}
	\end{minipage} & \begin{minipage}[b]{\linewidth}\raggedright
		aRMSE
	\end{minipage} \\
	\midrule\noalign{}
	\endhead
	\bottomrule\noalign{}
	\endlastfoot
	None (Raw signal) & 0.1 & 41.751 & 414.582 & 0.881 & -0.152 & 525.850 \\
	& 0.2 & 42.843 & 436.437 & 0.883 & -0.189 & 550.640 \\
	& 0.3 & 43.658 & 458.747 & 0.883 & -0.235 & 575.603 \\
	& 0.4 & 44.857 & 486.883 & 0.878 & -0.283 & 607.417 \\
	& 0.5 & 46.270 & 521.516 & 0.871 & -0.347 & 646.247 \\
	& 0.6 & 45.979 & 533.456 & 0.846 & -0.317 & 711.043 \\
	& 0.7 & 46.024 & 568.866 & 0.819 & -0.357 & 821.714 \\
	& 0.8 & 47.897 & 666.720 & 0.775 & -0.498 & 855.896 \\
	& 0.9 & 42.335 & 680.836 & 0.649 & -0.346 & 662.343 \\
	\(\mu\) & & 44.623 & 529.782 & 0.832 & -0.302 & 661.861 \\ \midrule\noalign{}
	DWT-Haar & 0.1 & 33.090 & 316.129 & 0.794 & 0.228 & 424.638 \\
	& 0.2 & 32.469 & 321.536 & 0.782 & 0.246 & 433.481 \\
	& 0.3 & 33.926 & 336.406 & 0.783 & 0.216 & 449.439 \\
	& 0.4 & 35.057 & 355.232 & 0.790 & 0.182 & 471.089 \\
	& 0.5 & 34.308 & 360.251 & 0.779 & 0.210 & 481.722 \\
	& 0.6 & 33.734 & 363.310 & 0.767 & 0.254 & 490.736 \\
	& 0.7 & 35.168 & 427.749 & 0.747 & 0.065 & 568.721 \\
	& 0.8 & 35.851 & 451.956 & 0.721 & 0.158 & 607.841 \\
	& 0.9 & 32.192 & 461.833 & 0.679 & 0.226 & 623.731 \\
	\(\mu\) & & 33.977 & 374.933 & 0.760 & 0.198 & 505.711 \\ \midrule\noalign{}
	Legendre & 0.1 & 33.365 & 315.387 & 0.801 & 0.232 & 423.325 \\
	& 0.2 & 33.645 & 324.701 & 0.803 & 0.231 & 435.620 \\
	& 0.3 & 33.467 & 331.136 & 0.791 & 0.240 & 444.985 \\
	& 0.4 & 33.903 & 344.888 & 0.796 & 0.229 & 463.019 \\
	& 0.5 & 35.901 & 376.203 & 0.796 & 0.138 & 496.640 \\
	& 0.6 & 34.501 & 373.463 & 0.784 & 0.212 & 501.772 \\
	& 0.7 & 34.614 & 402.475 & 0.750 & 0.172 & 536.007 \\
	& 0.8 & 35.219 & 443.336 & 0.738 & 0.190 & 596.664 \\
	& 0.9 & 32.241 & 481.656 & 0.690 & 0.158 & 645.646 \\
	\(\mu\) & & 34.095 & 377.027 & 0.772 & 0.200 & 504.853 \\ \midrule\noalign{}
	DCT & 0.1 & 32.877 & 313.104 & 0.795 & 0.243 & 420.794 \\
	& 0.2 & 32.823 & 321.144 & 0.792 & 0.248 & 432.058 \\
	& 0.3 & 33.173 & 326.694 & 0.788 & 0.260 & 439.919 \\
	& 0.4 & 33.404 & 341.165 & 0.779 & 0.245 & 458.309 \\
	& 0.5 & 34.383 & 356.587 & 0.783 & 0.226 & 478.401 \\
	& 0.6 & 33.597 & 358.875 & 0.768 & 0.272 & 486.128 \\
	& 0.7 & 34.001 & 404.248 & 0.748 & 0.164 & 542.202 \\
	& 0.8 & 35.156 & 437.848 & 0.728 & 0.210 & 591.559 \\
	& 0.9 & 32.741 & 468.723 & 0.686 & 0.202 & 663.260 \\ 
	\(\mu\) & & 33.572 & 369.820 & 0.763 & 0.230 & 501.403 \\
\end{longtable}

\end{document}